# Metasurface-on-Lithium-Niobate Architecture for Spin-Decoupled Second Harmonic Wavefront Shaping

Yiwen Liu[1], Chao Meng,[1,*] Sergey I. Bozhevolnyi[1], Fei Ding,[1,2,*]

[1]Centre for Nano Optics, University of Southern Denmark, Campusvej 55, Odense DK-5230, Denmark

[2]School of Electronic Science and Technology, Eastern Institute of Technology, Ningbo, China

*Correspondence: Chao Meng (chao@mci.sdu.dk) and Fei Ding (fding@eitech.edu.cn)

## Abstract

Nonlinear metasurfaces (NMSs) enable simultaneous generation and manipulation of harmonic light with ultrathin nanophotonic platforms. However, conventional NMSs typically adopt a monolithic single-layer configuration in which frequency conversion, polarization selection and wavefront engineering are intertwined to simultaneously occur within the same nonlinear meta-atoms, making these functionalities intrinsically interdependent and imposing constraints on material choice, structural symmetry, and the available geometric parameter space. Here, we propose and experimentally demonstrate a metasurface-on-lithium-niobate (MS-on-LN) architecture that physically separates harmonic light generation and wavefront manipulation. In this platform, a z-cut thin-film LN layer serves as a spin-selective second-harmonic source through its intrinsic bulk second-order nonlinearity, while an overlaid dielectric metasurface subsequently and independently molds wavefronts of generated harmonic fields. The two-step process, by virtue of harmonic generation being decoupled from wavefront shaping, removes symmetry constraints associated with conventional nonlinear meta-atoms and enables the direct transfer of well-established linear metasurface design strategies to harmonic wavefront engineering. As proof-of-concept demonstrations, we realize spin-decoupled second-harmonic beam steering and spin-multiplexed holography, achieving high-contrast routing of harmonic signals with orthogonal spin states into distinct diffraction channels and high-fidelity reconstruction of independent holographic images. By redirecting the harmonic wavefront control from the meta-atom engineering to the modular system-level design, our approach establishes a versatile route toward integrated nonlinear meta-optics and provides a scalable platform for advanced optical functionalities, including complex-amplitude modulation, vectorial field synthesis, structured-light generation, and nonlinear information processing.

## Introduction

Nonlinear metasurfaces (NMSs) provide a compact platform for manipulating nonlinear light–matter interactions at the nanoscale[1,2]. In conventional bulk nonlinear crystals, harmonic generation is mainly governed by the intrinsic material susceptibility and the propagation response accumulated over macroscopic interaction lengths[3], making subwavelength, pixel-level wavefront control difficult to implement. By contrast, NMSs employ spatially arranged subwavelength meta-atoms whose geometry, orientation, and symmetry can be engineered to tailor the generated harmonic fields[4,5]. Such subwavelength engineering enables precise harmonic wavefront shaping within an ultrathin integrated platform, leading to compact and versatile solutions for functionalities such as beam steering[6], focusing[7,8], holography[9,10], and orbital-angular-momentum generation[11,12].

In wavefront-shaping NMSs, circular polarization (CP)-based approaches are particularly attractive because they naturally combine nonlinear optical selection rules with geometric-phase engineering [10,13–16]. In such single-layer NMSs (Fig. 1A), the connection between nonlinear optical selection rules and geometric-phase engineering is implemented at the meta-atom level: the symmetry of the nonlinear meta-atom determines the allowed CP channel of the generated harmonic field, while its in-plane orientation controls the phase imparted to that channel. The channel-selection aspect can be understood through nonlinear optical selection rules[17,18], which were first established for natural nonlinear crystals and later applied to artificial meta-atoms[19]. For a circularly polarized fundamental wave (FW) propagating along an $n$-fold rotational symmetry axis, the allowed harmonic orders satisfy $m = nl \pm 1$, where $l$ is an integer and $n$ denotes the rotational symmetry order of the system; the "+" and "–" signs correspond to harmonic generation with the same and opposite CP handedness relative to the incident field, respectively[14,20]. Meanwhile, nonlinear geometric phase provides a route for continuous phase modulation: through spin–rotation coupling, the in-plane rotation of meta-atoms induces a phase shift in the nonlinear polarization, whose slope with respect to the rotation angle is determined by the harmonic order and the spin conversion channel. Consequently, the rotational symmetry and in-plane orientation of the meta-atom jointly determine the generated spin channel and the phase profile of the harmonic wave, establishing a symmetry–orientation framework for spin-dependent harmonic wavefront engineering.

As noted above, in such single-layer NMSs, harmonic generation, spin-channel selection, and phase engineering are all encoded within the same meta-atom design (Figure 1A). Although compact, this approach introduces mutual coupling among multiple design considerations, whereby optimizing one functionality often affects the others, resulting in inevitable trade-offs among conversion efficiency, polarization purity, amplitude uniformity, and phase controllability. Taking second-harmonic (SH) NMSs as an example, achieving a single circularly polarized SH channel under CP excitation typically requires at least $C_3$ rotational symmetry[18], with recent studies further highlighting the importance of both nanostructure geometry and lattice symmetry[21]. To realize spin-decoupled SH phase control, additional phase degrees of freedom beyond the rotation-induced geometric phase of each meta-atom, such as propagation or resonant phases, must be introduced. These mechanisms often rely on modifying the meta-atom dimensions or resonance conditions, thereby compromising the uniformity of the SH amplitude distribution. Moreover, the effective interaction length for SH generation is generally confined to the localized near-field region of the meta-atoms, which is typically only tens of nanometers for plasmonic structures and therefore limits the scope for improving the SHG efficiency. While simultaneous resonances at both the FW and SH wavelengths can effectively enhance the conversion efficiency[22,23], their implementation adds considerable design complexity and imposes additional constraints on design flexibility. Therefore, although single-layer NMSs provide a compact platform for harmonic wave generation and manipulation, the associated functionalities remain intrinsically intertwined within the same nonlinear meta-atom, hindering the independent and flexible control of harmonic wave generation, polarization, phase, and conversion efficiency.

Here, rather than implementing all functionalities within a single nonlinear meta-atom, we revisit nonlinear selection rules from the material perspective and propose a metasurface-on-lithium niobate (MS-on-LN) architecture for spin-decoupled SH wavefront control, as illustrated in Fig. 1B. In this design, SH generation and spin-channel selection are governed by the intrinsic bulk $\chi^{(2)}$ response of the z-cut LN layer, whereas the metasurface nanostructure, fabricated directly on top of a thin-film LN layer, can be designed with arbitrary geometries to independently tailor the wavefront of the generated SH field. This separation effectively decouples nonlinear frequency conversion from wavefront manipulation, thereby substantially relaxing the symmetry and design constraints imposed on conventional nonlinear meta-atoms. Specifically, we employ birefringent titanium dioxide ($TiO_2$) nanopillars that combine geometric and propagation phases to achieve spin-decoupled SH wavefront modulation at FW and SH wavelengths of approximately 1550 nm and 775 nm, respectively. We experimentally demonstrate three representative devices: an MS-on-LN grating for spin-dependent SH beam steering, and two spin-multiplexed SH meta-holograms for reconstructing image and letter patterns. The MS-on-LN grating exhibits high-contrast spin-dependent SH beam steering, in which opposite FW CPs direct the generated SH signals into the designed $m = -1$ and $m = 2$ diffraction orders, accounting for 89.3% and 80.9% of the total measured SH intensity (97.2% and 88.8% in simulation), respectively. The MS-on-LN meta-holograms enable spin-multiplexed reconstruction of two independent images in orthogonal SH CP channels, demonstrating high-fidelity and sharp image quality for both photographic and letter-pattern holograms. Beyond these demonstrations, the proposed MS-on-LN architecture provides a general route for transferring mature wavefront-engineering strategies developed for linear metasurfaces to nonlinear optical signals, thereby expanding the design space of nonlinear metasurfaces and enabling advanced functionalities that are difficult to realize using conventional nonlinear meta-atom designs.

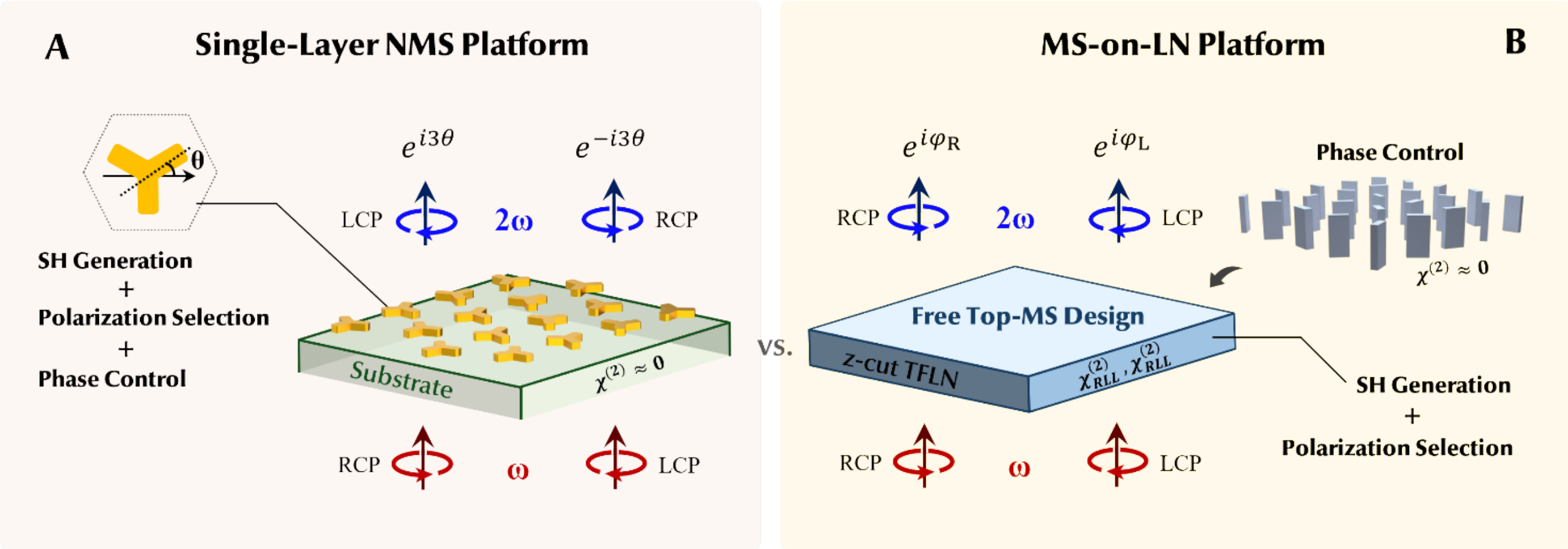

**Fig. 1. Comparison of conventional single-layer NMS and MS-on-LN architectures for Second-Harmonic wavefront control.** (A) Conventional single-layer NMS platform based on centrosymmetric materials lacking bulk $\chi^{(2)}$. In this configuration, SH generation, spin-channel selection, and phase control are simultaneously implemented within the same nonlinear meta-atom. Circularly polarized SH generation typically requires at least $C_3$ rotational symmetry, imposing significant constraints on the meta-atom geometry and limiting the design flexibility of the NMS. (B) Proposed MS-on-LN configuration. SH generation and spin-channel selection are governed by the intrinsic bulk $\chi^{(2)}$ response of the z-cut LN layer, whereas the overlaid metasurface is designed exclusively for wavefront engineering. This separation effectively decouples nonlinear frequency conversion from phase control, relaxing symmetry constraints on the meta-atoms and enabling greater flexibility for harmonic wavefront engineering.

# 2 Results and Discussion

## 2.1 Second-Order Susceptibility of z-cut LN in the CP Basis

LN is a trigonal nonlinear crystal belonging to the 3m ($C_{3v}$) point group. The form of the second-order susceptibility tensor is determined by the crystal symmetry and the chosen coordinate system. Here, we consider a normally incident FW propagating along the *z*-axis, with the electric field polarized in the transverse *x*-*y* plane. Under this condition, the in-plane SH nonlinear polarization can be expressed in the linear polarization basis and then transformed into the CP basis (see details in Supplementary Materials S1.1) as

$$P_{l_1}^{2\omega} = \varepsilon_0 \sum_{l_2,l_3} \chi_{l_1 l_2 l_3}^{(2)} E_{l_2}^{\omega} E_{l_3}^{\omega}, \quad \{l_1,l_2,l_3\} \epsilon \{x,y\}$$
$$P_{c_1}^{2\omega} = \varepsilon_0 \sum_{c_2,c_3} \chi_{c_1 c_2 c_3}^{(2)} E_{c_2}^{\omega} E_{c_3}^{\omega}, \quad \{c_1,c_2,c_3\} \epsilon \{R,L\} \tag{1}$$

where $l_i \in \{x, y\}$ and $c_i \in \{R, L\}$ denote the linear-polarization and CP components, respectively. For a z-cut LN film under normal-incidence excitation, only the in-plane electric-field and nonlinear-polarization components contribute to the transmitted SH signal. In the linear polarization basis, the nonlinear response can therefore be simplified as

$$P_x^{2\omega} = 2\chi_{xxy}^{(2)} E_x^{\omega} E_y^{\omega}$$
$$P_y^{2\omega} = \chi_{yxx}^{(2)} E_x^{\omega} E_x^{\omega} + \chi_{yyy}^{(2)} E_y^{\omega} E_y^{\omega} \tag{2}$$

These expressions indicate that the SH nonlinear polarization involves coupled contributions from the orthogonal in-plane FW components. As a result, the polarization correspondence between the FW and SH fields is not directly revealed in the Cartesian basis. Transforming Eq. (2) into the CP basis yields

$$\begin{aligned} P_R^{2\omega} &= \chi_{RLL}^{(2)} E_L^{\omega} E_L^{\omega} \\ P_L^{2\omega} &= \chi_{LRR}^{(2)} E_R^{\omega} E_R^{\omega} \end{aligned} \tag{3}$$

Equation (3) reveals that only the cross-polarized SH channels remain nonzero, indicating that the generated SH polarization has the opposite handedness to that of the incident FW in the CP basis. This behavior is a direct consequence of the intrinsic bulk $\chi^{(2)}$ tensor of z-cut LN and is independent of metasurface structuring. Experimental measurements on a bare 300-nm-thick z-cut LN substrate show comparable SH intensities and high cross-polarization purity under the two orthogonal CP FW excitations (Supplementary Materials S1.2). The measured FW–SH polarization mapping agrees well with the theoretical prediction, confirming that, under normal incidence, z-cut LN functions as an intrinsic spin-selective SH source.

### 2.2 Spin-Decoupled Metasurface Unit cell for SH Wavefront Engineering

As confirmed by both theory and experiment, z-cut LN layer acts as a spin-selective SH source, generating SH signals with CP handedness opposite to that of the FW under normal incidence. Based on this property, we propose a compact and flexible MS-on-LN architecture, in which a metasurface is designed and fabricated directly on top of the LN layer and is responsible solely for shaping the generated SH wavefront. This functional separation between SH generation and modulation substantially enhances the design flexibility of the metasurface, in contrast to conventional single-layer NMSs, where the relevant functionalities remain intrinsically coupled within the same nonlinear meta-atom (Fig. 1). In this architecture, SH generation and polarization selection are governed by the bulk $\chi^{(2)}$ response of the z-cut LN layer, providing an efficient and robust nonlinear source than the surface-induced nonlinearities typically exploited in conventional single-layer NMSs. The overlaid metasurface then serves as a post-processing layer that modulates the high-purity CP SH generated by the LN layer, allowing a wide range of design strategies developed for linear metasurfaces to be directly transferred to the SH regime. As a proof of concept, we employ birefringent rectangular nanopillar meta-atoms that combine propagation and geometric phases[24] to achieve independent SH phase control of two orthogonal SH CP channels.

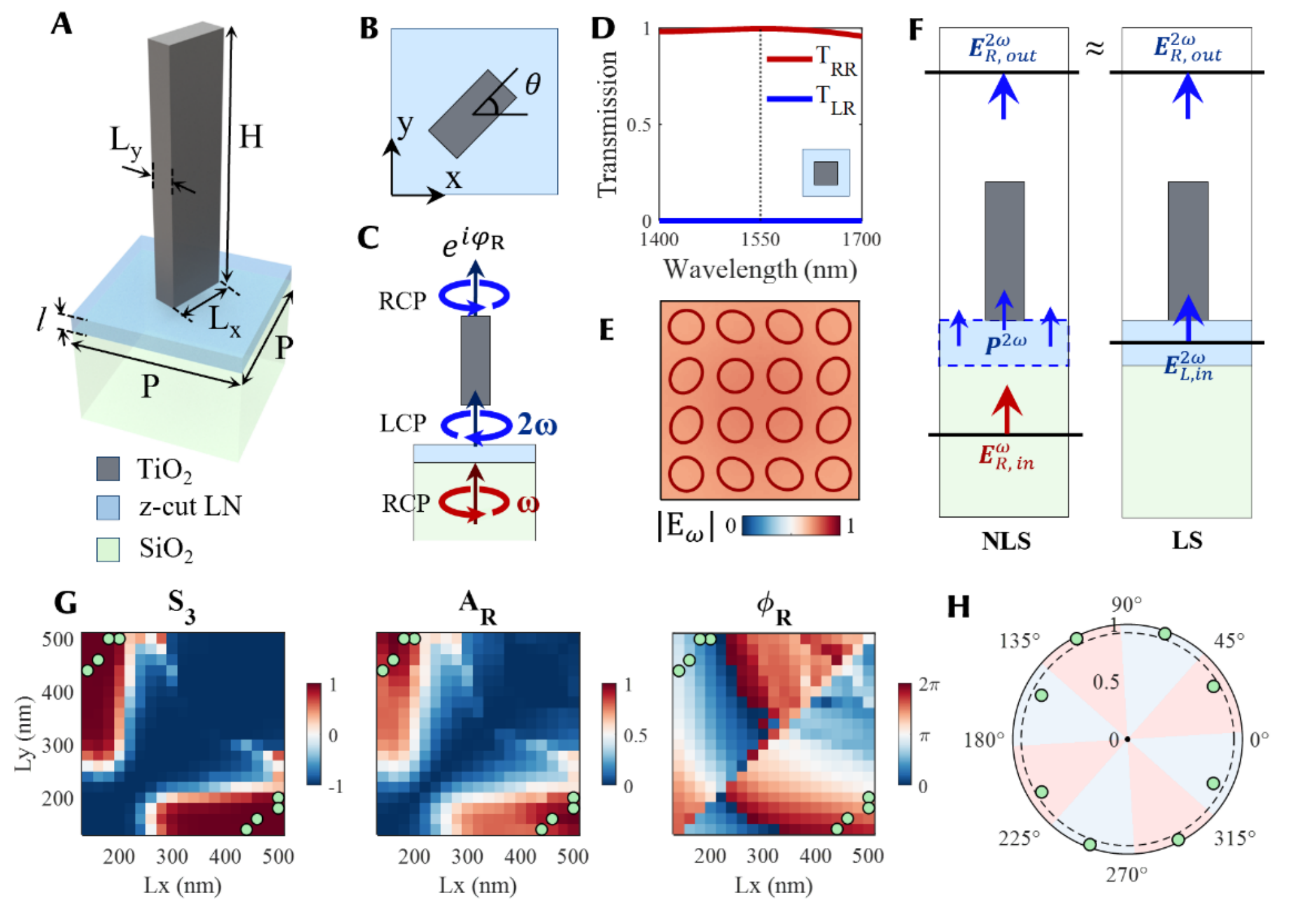

**Fig. 2. MS-on-LN unit-cell design for spin-decoupled SH phase control.** (A) Unit-cell geometry consisting of a $TiO_2$ nanopillar on a z-cut LN layer supported by a $SiO_2$ substrate. (B) Top view of the anisotropic nanopillar, where $\theta$ denotes the in-plane rotation angle. (C) Schematic of polarization and phase evolution along the propagation direction. The LN layer generates cross-polarized SH, which is subsequently polarization-converted and phase-modulated by the $TiO_2$ nanopillar. (D) Calculated transmission spectrum of a representative unit cell ($L_x = L_y = 320$ nm) under RCP FW incidence, showing high transmittance and no resonance near the FW wavelength of interest (1550 nm). (E) Electric-field amplitude and local polarization state at the FW wavelength at the mid-plane of the LN layer across the entire unit-cell area of the representative unit cell shown in (D). The FW field remains nearly uniform across the unit cell and preserves its RCP state within the LN layer. (F) Comparison of the nonlinear simulation (NLS) and linear simulation (LS) schemes used in this work, demonstrating that the nonlinear response can be approximated by the corresponding linear response at the SH wavelength. (G) Calculated maps of SH polarization purity $S_3$, normalized RCP SH amplitude $A_R$ (relative to bare LN), and SH phase $\varphi_R$ as functions of $L_x$ and $L_y$ under RCP FW illumination at 1550 nm wavelength, obtained from NLS scheme. The selected meta-atoms are marked by green circles. (H) Complex amplitudes of the selected eight meta-atoms, where the radial coordinate represents the normalized SH amplitude $A_R$ (the dashed circles corresponding to $A_R$ = 0.93) and the angular coordinate denotes the SH phase from 0–2$\pi$.

To realize spin-decoupled phase response for SH waves, we designed a series of different birefringent meta-atoms within square unit cell lattices, as shown in Fig. 2. These meta-atoms operate as nano-half-wave plates at the SH wavelength, enabling efficient polarization conversion and SH phase modulation of the generated SH waves. As shown in Figure 2(A), the unit cell consists of a $TiO_2$ nanopillar placed on a LN layer, supported by a $SiO_2$ substrate. The thickness of the LN is t = 300 nm. The $TiO_2$ nanopillar has a height of H = 800 nm and in-plane dimensions $L_x$ and $L_y$, arranged in a square lattice with a period of P = 600 nm. The design FW and SH wavelengths are 1550 nm and 775 nm, respectively. The unit-cell design follows the well-established framework for independent phase control of orthogonal polarization states based on linearly birefringent elements[24]. Under CP illumination, the meta-atoms are designed to operate close to a half-wave plate condition, enabling polarization conversion between the two spin channels[25]. Its phase response consists of a propagation phase determined by the structural birefringence and a geometric phase introduced by the in-plane rotation angle $\theta$ [Fig. 2(B)]. These two degrees of freedom enable independent control of the phase responses of the two CP channels. As illustrated in Fig. 2(C), the LN layer first generates an SH field with handedness opposite to that of the incident FW. The $TiO_2$ nanopillar then acts as a nano-half-wave plate at the SH wavelength, converting the generated SH field back to the same spin channel as the incident FW while imparting desired phase profile. Therefore, the LN layer governs the nonlinear spin-selection process, whereas the nanopillar controls the SH phase response, providing a direct physical implementation of the decoupled MS-on-LN architecture.

In the unit-cell design, the lattice period is deliberately chosen based on two key considerations. First, to suppress higher-order diffraction in transmission, the period must remain subwavelength with respect to the SH wavelength. Second, the period is chosen such that no guided-mode resonance is excited near the FW wavelength, thereby avoiding field redistribution inside the LN layer that would alter the polarization mapping between FW and SH. Figure 2(D) shows the transmission spectrum of a representative case with $L_x = L_y = 320$ nm and P = 600 nm near the FW wavelength range under RCP incidence. The transmission remains high, and no resonance feature is observed around the FW wavelength. As shown in Fig. 2(E), the FW electric field exhibits a nearly uniform amplitude distribution at the mid-plane of the LN layer, while the local polarization state, indicated by the ellipses, remains circular throughout the unit cell. These results indicate that the $TiO_2$ nanopillars only weakly perturb the FW field inside the LN layer, allowing the LN to retain its intrinsic spin-selective SH generation characteristics. This observation supports the theoretical prediction presented in Section 2.1 that the LN layer governs SH generation and polarization selection, while the top $TiO_2$ nanopillars act as an independent layer for SH wavefront engineering. Under this cascaded framework, the nonlinear response of the unit cell can be approximated by the corresponding linear response at the SH wavelength, as shown in Fig. 2(F). This approximation reflects the physical separation of harmonic generation and wavefront manipulation in the MS-on-LN architecture and is validated by the good agreement between the nonlinear simulation (NLS) results obtained under FW incidence (left panel in Fig. 2(F)), and the linear simulation (LS) results obtained under direct SH wave incidence (right panel in Fig. 2F) (see Supplementary Section S2 for details). Besides substantially reducing the computational cost of unit-cell optimization, this approach also enables pre-characterization of the fabricated samples using a linear light source at the SH wavelength.

Following the spin-decoupled phase design strategy, a set of nanopillars with different dimensions is selected such that they operate as half-wave plates (HWPs) with phase covering a 0–2π phase range. As shown in Fig. 2(G), the Stokes parameter $S_3$, amplitude and phase of the SH wave after the MS-on-LN unit cell are calculated as functions of $L_x$ and $L_y$ under RCP FW incidence ($S_3 = 1$). Eight nanopillars are selected based on three criteria: $S_3 > 0.92$, normalized amplitude of the co-polarization SH component, $A_R > 0.8$, and phase $\phi_R$ distributed relatively uniform over the full 0–2π range. The selected unit cell designs are summarized in Fig. 2(H), where the radial coordinate represents the normalized SH amplitude relative to the bare LN layer and the angular coordinate denotes the SH phase. The selected unit cells exhibit SH amplitudes that remain close to unity, with an average normalized amplitude of 0.93, indicating that the phase-encoded SH generation maintains nearly the same amplitude as that of the bare LN film. The corresponding geometrical parameters and simulated responses of the selected eight unit cell designs are listed in Supplementary Materials Table S1. In the following sections, these unit cells are employed as the fundamental building blocks for spin-selective SH beam steering and spin-multiplexed SH holography, demonstrating the versatility of the proposed MS-on-LN platform for harmonic wavefront engineering.

### 2.3 Spin-Dependent SH Beam Steering

To demonstrate the spin-decoupled wavefront engineering capability of the proposed platform, we first designed an MS-on-LN blazed grating for spin-dependent SH beam steering. In this design, the two CP SH channels are independently encoded with distinct periodic linear phase gradients, enabling the generated SH signals to be directed into different diffraction orders. As illustrated in Fig. 3(A), under LCP FW incidence, the device emits an LCP SH signal directed into the $m = -1$ diffraction order, whereas under RCP FW incidence, it emits an RCP SH signal directed into the $m = +2$ order. Accordingly, the MS-on-LN grating is constructed from an eight-element supercell (Fig. 3B), which is periodically repeated along the $x$ and $y$ directions with $\Lambda_x = 4.8$ μm and $\Lambda_y = 0.6$ μm. The designed discrete linear phase profiles $\varphi_R$ undergoes two full 0–$2\pi$ cycles within the supercell, while $\varphi_L$ completes a single cycle with a phase gradient of opposite sign, corresponding to the distinct blazed diffraction orders assigned to the two CP SH channels.

The designed MS-on-LN gratings were fabricated using electron-beam lithography (EBL), thin-film deposition, and inductively coupled plasma (ICP) etching processes, as detailed in Supplementary Section S3. The grating has an overall size of 96 μm × 96 μm. Along the x-direction, it contains 20 grating supercells, each consisting of eight unit cells, while along the y-direction the unit cells are periodically repeated to give the same lateral size. A scanning electron microscopy (SEM) image of the fabricated grating (Fig. 3(C)) shows good agreement with the designed supercell structure shown in Fig. 3(B).

To characterize the MS-on-LN grating, a fiber-coupled femtosecond (fs) laser (ARCHEv3, FYLA Laser) with a central wavelength of 1550 nm, a pulse duration of approximately 200 fs, and a repetition rate of 65 MHz was used as the excitation source. The laser beam was converted into CP FW using polarization optics, and then weakly focused onto the grating. Details of the experimental setup are provided in Supplementary Section S4. Fig. 3(D) shows the far-field diffraction patterns of the generated SH signals under circularly polarized FW incidence. The top row shows the far-field SH signal generated from the bare LN substrate without the $TiO_2$ metasurface, where the 0th-order spot serves as both a positional reference and an intensity normalization baseline for the relative efficiency estimation in Fig. 3(E). Here, the notation R→R denotes the corresponding polarization states of the input FW and output SH signals, respectively. For the bare LN substrate, the generated SH polarization exhibits opposite handedness to that of the incident FW, and is therefore measured in the R→L channel. The bottom two rows correspond to the MS-on-LN grating under orthogonal CP FW excitation. As designed, the SH signals are deflected into the $m = -1$ and $m = +2$ diffraction orders under LCP and RCP FW illumination, respectively. A high intensity contrast is observed between the desired and undesired diffraction orders, with the latter being nearly imperceptible in the measured far-field patterns.

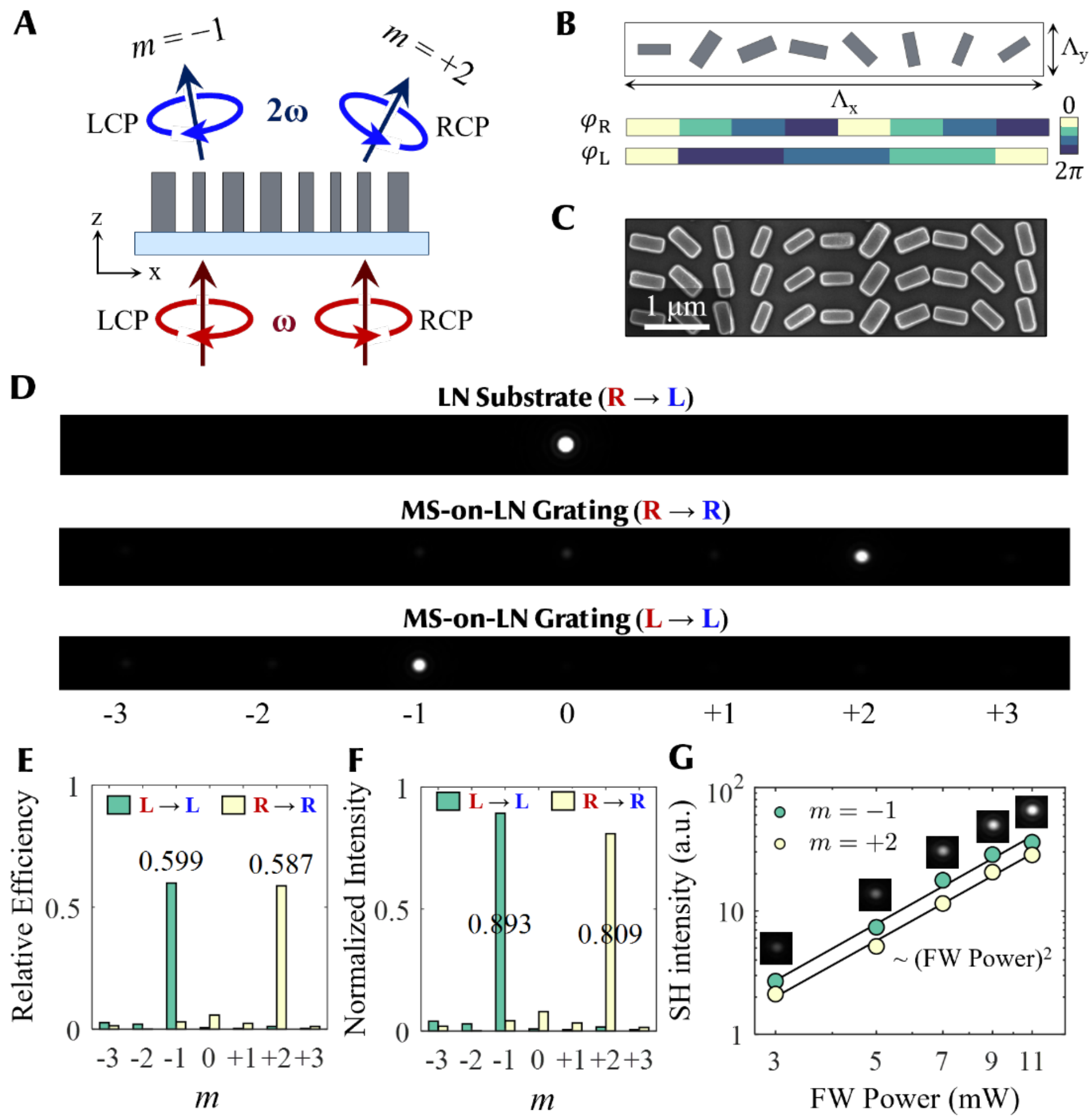

**Fig. 3. Spin-decoupled SH beam steering with a MS-on-LN grating.** (A) Schematic illustration of spin-dependent SH beam steering, where orthogonal circularly polarized FWs generate SH signals with the same handedness and are deflected into different diffraction orders. (B) Supercell design and the corresponding phase distributions independently encoded for the two SH CP channels. (C) Scanning electron microscopy (SEM) image of a representative region of the fabricated MS-on-LN grating. (D) Measured SH far-field diffraction patterns. The top row shows the SH response from a bare z-cut LN substrate, while the bottom two rows correspond to the MS-on-LN grating under orthogonal CP FW illumination, exhibiting SH beam deflection into the designed $m = -1$ and $m = +2$ diffraction orders, respectively. Here, the notation R→R enotes the polarization states of the incident FW and the generated SH signals, respectively. (E) Measured relative SH diffraction efficiencies for different diffraction orders under orthogonal CP FW incidence, obtained from the SH intensities of the MS-on-LN grating in the corresponding SH polarization channels and normalized to the SH intensity generated by a bare z-cut LN substrate in the same SH polarization channel. (F) Measured SH intensities at the designed diffraction orders, normalized by the total intensity summed over the displayed orders. (G) Measured SH intensities at the designed diffraction orders as a function of FW power, exhibiting the expected quadratic scaling dependence characteristic of SH generation.

To quantify the performance of the MS-on-LN grating, we evaluate the relative SH diffraction efficiency and diffraction contrast from the recorded far-field diffraction patterns. The relative SH diffraction efficiency is defined as the intensity of a given diffraction order normalized to the generated SH intensity from a bare LN reference, while the diffraction contrast is defined as the intensity of that diffraction order normalized to the total SH intensity summed over diffraction orders $|m| \leq 3$. As shown in Fig. 3(E), the measured relative SH diffraction efficiencies at the designed diffraction orders $m = -1$ and $m = +2$ are 0.599 and 0.587, respectively, compared with the simulated values of 0.881 and 0.747 shown in Figs. S6(C, D). The lower experimental efficiencies can be attributed to fabrication imperfections and non-ideal plane-wave excitation, which may redistribute power into undesired diffraction orders and/or polarization channels. Despite the reduced relative efficiencies, the measured diffraction contrast remains high. As shown in Fig. 3(F), the designed diffraction orders dominate the measured far-field patterns, yielding diffraction contrasts of 0.893 ($m = -1$, LCP FW) and 0.809 ($m = +2$, RCP FW), respectively, comparable to the simulated values of 0.972 and 0.888 shown in Fig. S7(A). These results verify that the generated SH power is predominantly concentrated into the designed diffraction channels, demonstrating efficient spin-dependent SH wavefront engineering enabled by the MS-on-LN architecture. Finally, we examine the power dependence of the detected SH intensity as a function of the pump power to confirm its nonlinear origin. As expected, the SH intensity integrated over the designed diffraction orders scales quadratically with the incident FW power [Fig. 3(G)].

**2.4 Spin-Multiplexed SH Meta-Holography**

In addition to the MS-on-LN grating, which validates the spin-decoupled wavefront-engineering capability of the platform, we further demonstrate its versatility through the development of spin-multiplexed SH meta-holography. Compared with SH beam steering using a grating, holography requires spatially resolved phase control at the two-dimensional (2D) pixel level, thereby providing a more stringent test of the SH wavefront engineering capability of the MS-on-LN architecture. In this section, we develop spin-multiplexed MS-on-LN SH meta-holograms capable of generating distinct SH holographic images under orthogonal circularly polarized FW illumination conditions, as illustrated in Fig. 4(A). Given that our design yields SH signals with nearly uniform amplitude, we adopt a phase-only holography strategy, where the amplitude of each pixel is fixed and the image reconstruction is achieved solely through spatial phase modulation. The phase profiles are retrieved using an stochastic gradient descent (SGD)-based algorithm, following a recent study[26] that demonstrated improved hologram performance compared with the Gerchberg–Saxton algorithm. Details of the target holographic images, hologram design process, phase quantization, and beam-size dependence are provided in Supplementary Section S6. More generally, the proposed MS-on-LN platform is compatible with established complex-amplitude design strategies from linear metasurfaces[27,28], where the geometry and dimensions of the meta-atoms are further tailored to modulate both the amplitude and phase of the generated SH field.

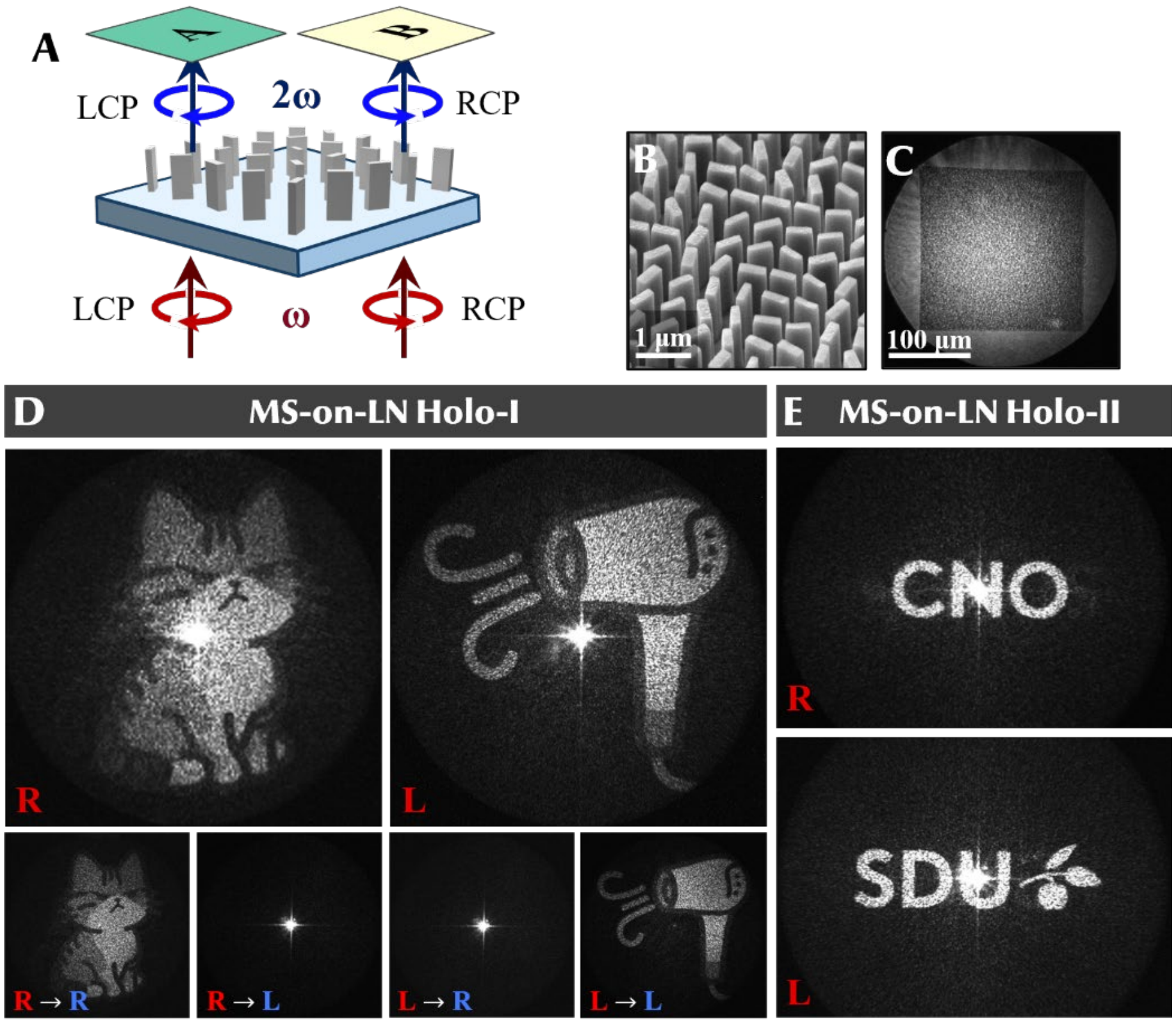

**Fig. 4. Spin-multiplexed SH meta-holography.** (A) Schematic illustration of spin-multiplexed meta-holography based on the MS-on-LN configuration, where orthogonal CP FW excitations generate SH signals with the same handness and reconstruct distinct holographic images in the far field. (B) SEM image of part of the fabricated MS-on-LN Holo-I. (C) Real-image-plane image showing the metasurface region selected by a circular iris, which defines the collection window for the far-field SH holographic measurements. The observed signal corresponds to the generated SH emission at the MS plane. (D) Measured holographic images generated by MS-on-LN Holo-I and captured at the Fourier plane under orthogonal CP FW illumination. The top row shows the results obtained without polarization filtering in the detection path, where two distinct high-quality images (an angry cat and a hair dryer) are reconstructed with negligible crosstalk. The bottom row shows the corresponding results obtained with CP filtering in the detection path, where the 0th-order SH spots are predominantly observed in the cross-polarized channel, yielding cleaner SH holographic images in the co-polarized channel. (E) Measured holographic images generated by MS-on-LN Holo-II at the Fourier plane under orthogonal CP FW illumination, without polarization filtering in the detection path, demonstrating high-fidelity reconstruction of two distinct letter patterns ("CNO" and "SDU").

We designed two MS-on-LN meta-holograms, denoted as MS-on-LN Holo-I and II. Each metasurface consists of 333 × 333 meta-atoms with an overall size of 199.8 μm × 199.8 μm, as shown in Fig. 4B and C. MS-on-LN Holo-I encodes two photographic images, while MS-on-LN Holo-II encodes two letter patterns. To characterize the MS-on-LN meta-holograms, a weakly focused fs laser beam with a diameter of approximately 400 μm is used to illuminate the metasurface, which is significantly larger than the metasurface size to approximate uniform plane-wave excitation. The metasurface profile and the generated SH signals can be observed simultaneously in the real image plane using a CMOS camera, as shown in Fig. 4(C). In the measurements, an iris is used as a spatial filter to define a collection window that fully contains the square metasurface area, thereby avoiding truncation of the reconstructed phase profile. Inevitably, SH signals generated by the surrounding LN substrate outside the metasurface are also collected, contributing to a residual 0th-order spot. However, because these signals do not undergo polarization conversion, they can be selectively suppressed using a CP analyzer, as shown in Fig. 4(D).

Under orthogonal CP FW illumination, spin-multiplexed SH holographic images are reconstructed and observed at the Fourier plane, as shown for MS-on-LN Holo-I in Figure 4(D), corresponding to an angry cat and a hair dryer. The top row presents the results without polarization filtering of the SH signals, where both images are clearly visible under orthogonal CP FW excitations. The absence of noticeable crosstalk between the two channels indicates that the generated SH signals maintain well-defined CP channels. A prominent 0th-order spot is observed in each image, which mainly originates from the SH signal generated in the surrounding LN substrate that is not blocked by the iris. When polarization selection is applied to the SH signals (bottom row), the 0th-order spot is predominantly filtered into the co-polarized channel, leaving a cleaner holographic image in the cross-polarized channel. Furthermore, MS-on-LN Holo-II also exhibits high-quality spin-multiplexed SH holographic reconstruction. Fig. 4(E) shows the reconstructed two holographic letter patterns ("CNO" and "SDU") without SH polarization filtering, featuring well-defined edges and high intensity contrast. Polarization resolved measurements (Fig. S12) further confirm that the holographic images are generated in the co-polarized channel, while the $0^{th}$-order spots are generated in the cross-polarized channel, consistent with the behavior observed for MS-on-LN Holo-I. Together, these results validate the proposed MS-on-LN configuration as a versatile platform for spin-multiplexed SH meta-holography, capable of reconstructing diverse target images with high fidelity and high polarization selectivity.

## 3 Conclusion

In summary, we have introduced and experimentally demonstrated an MS-on-LN platform that enables spin-selective SH generation and spin-decoupled SH wavefront control. Unlike conventional single-layer NMSs, where harmonic generation, polarization selection, and phase modulation are intrinsically intertwined within the same meta-atom, leading to restricted design flexibility and/or compromised optical performance, our approach decouples these roles into two functional layers. Specifically, the z-cut LN provides the intrinsic $\chi^{(2)}$ response and defines the FW–SH spin-conversion pathway, while the metasurface nanostructures provide a versatile and highly flexible layer for tailoring the generated SH wavefront. Based on the MS-on-LN platform, we experimentally validated this concept through a grating for spin-dependent SH beam steering and two meta-holograms for spin-multiplexed SH holography. Under orthogonal circularly polarized FW excitations, the grating enables high-contrast routing of SH signals into distinct diffraction channels, while the meta-holograms reconstruct independent holographic images with high fidelity and low crosstalk, validating the capability of the proposed platform for two-dimensional pixel-level control of harmonic optical wavefronts.

While the present work focuses on demonstrating the functional separation of SH generation and wavefront manipulation, conversion efficiency remains an important consideration for practical nonlinear photonic devices. As discussed in Supplementary Section S8, increasing the LN-layer thickness toward the coherence length provides a straightforward strategy for extending the nonlinear interaction length and enhancing SHG efficiency without altering the general metasurface design concept of the MS-on-LN architecture. Such optimization would improve the signal-to-noise ratio of nonlinear beam steering and holographic reconstruction while reducing the required pump power, both of which are essential for practical implementations.

More broadly, the proposed architecture transforms the control of harmonic wavefronts from a meta-atom-level optimization problem into a system-level design framework. Because the metasurface layer is no longer required to provide the nonlinear response, the constraints on meta-atom geometry, material choice, and fabrication route are substantially relaxed, enabling the implementation using a broader range of linear optical platforms, including recently emerging three-dimensionally printed polymer micro/nanostructures[28,29]. More importantly, well-established wavefront-engineering strategies developed for linear metasurfaces can be directly transferred to nonlinear optical signals without being constrained by nonlinear selection rules or symmetry requirements. This transferability extends the platform beyond the phase-only demonstrations presented here and enables compatibility with advanced linear metasurface concepts, including complex-amplitude modulation[28,30], vectorial polarization shaping[31], spin-controlled SH vortex generation, optical skyrmion generation[32], and beyond. We therefore envision this architecture as a scalable and versatile platform for integrated nonlinear meta-optics.

# 4 Methods

## 4.1 Numerical Simulation

All simulations in this work were performed using the Electromagnetic Waves, Frequency Domain (EWFD) module of COMSOL Multiphysics 6.3, based on the finite-element method. The metasurface unit cell was modelled with periodic boundary conditions along the lateral directions and normally incident port excitation. The SHG process was simulated using a two-step frequency-domain approach. In the first step, a linear simulation was performed at the fundamental wavelength to obtain the local FW electric-field distribution. In the second step, the second-order nonlinear polarization $\boldsymbol{P}^{2\omega}$ induced within the LN layer was calculated from the simulated FW field and used as a source term in a separate frequency-domain simulation at the SH wavelength. The generated SH fields were then collected on a horizontal monitor plane in air above the unit cell, from which the SH Stokes parameter $S_3$, amplitude and phase were extracted for unit-cell selection. The SH amplitude was normalized to the corresponding SH amplitude generated from a bare z-cut LN substrate under identical excitation conditions.

## 4.2 Sample Fabrication

The metasurface samples were fabricated using electron-beam lithography (EBL), thin-film deposition, and inductively coupled plasma (ICP) etching. The samples were patterned on a commercially sourced z-cut LN substrate consisting of a 300-nm-thick LN bonded to a 2.5-μm-thick $SiO_2$ buffer layer on a 500-μm-thick quartz substrate. After substrate cleaning, an 800-nm-thick $TiO_2$ layer was deposited on the LN substrate using an optical thin-film deposition system. A 170-nm-thick layer of ZEP520A electron-beam resist was then spin-coated onto the $TiO_2$ surface and patterned by EBL with a beam current of 2 nA. After development, a 50-nm-thick Cr layer was deposited and transferred by lift-off to form a hard mask. The $TiO_2$ layer was subsequently etched to a depth of 800 nm by ICP etching, and the residual Cr mask was removed using a chromium etchant, leaving the fabricated $TiO_2$ nanopillar structures on the LN substrate. Additional fabrication details and SEM images of the fabricated samples are provided in Supplementary Section S3.

### 4.3 Optical Characterization

A fiber-coupled femtosecond laser (ARCHEv3, FYLA Laser) with a central wavelength of 1550 nm, a pulse duration of approximately 200 fs, and a repetition rate of 65 MHz was used as the FW excitation source. The incident beam was prepared as a weakly focused Gaussian beam with a diameter of approximately 400 μm on the sample plane and a typical average power of 20 mW. The generated SH signal was collected in transmission using a 40× objective (UPLFLN 40X, NA = 0.75), spectrally filtered to remove the residual FW, and imaged in either the Fourier plane or the real plane using a CMOS camera (ORCA-Flash4.0 LT, Hamamatsu). For polarization-resolved measurements, a quarter-wave plate and a linear polarizer were placed in the SH collection path to select the desired CP output channel. The complete optical setup and all optical components are shown in Supplementary Section S4.

## Acknowledgements

This work was supported by the National Natural Science Foundation of China (62575151), Zhejiang Provincial Natural Science Foundation of China (LR26F050003), Carlsberg Foundation (CF24-1777), Villum Fonden (37372 and 50343) and funding from the European Union (101185769–OPTIPATH). Y.L. conceived the idea and performed the simulations. C.M. and Y.L. built the experimental setup, and Y.L. carried out the measurements. C.M., F.D. and S.I.B. supervised the project. Y.L., C.M., F.D and S.I.B. analysed the data. F.D. and C.M. secured funding for the project. Y.L. drafted the manuscript. All authors discussed the results and contributed to the revision of the manuscript.

## Conflicts of Interest

The authors declare no conflict of interest.

## Data Availability Statement

The data that support the findings of this study are available from the corresponding author upon reasonable request.

Supplementary Materials for

# Metasurface-on-Lithium-Niobate Architecture for Spin-Decoupled Second Harmonic Wavefront Shaping

Yiwen Liu *et al.*

This document includes:

## Section S1. Spin-selective Second-Harmonic Generation in z-cut Lithium Niobate

### *S1.1 Derivation of the Second-Order Susceptibility Tensor in the Circular-Polarization Basis*

Lithium niobate (LN) is a trigonal nonlinear crystal belonging to the $3m(C_{3v})$ point group. In the Cartesian coordinate system, the second-order nonlinear response of LN can be written in tensor form as

$$\begin{bmatrix} P_x^{2\omega} \\ P_y^{2\omega} \\ P_z^{2\omega} \end{bmatrix} = \varepsilon_0 \begin{pmatrix} 0 & 0 & 0 & 0 & \chi_{xxz}^{(2)} & \chi_{xxy}^{(2)} \\ \chi_{yxx}^{(2)} & \chi_{yyy}^{(2)} & 0 & \chi_{yyz}^{(2)} & 0 & 0 \\ \chi_{zxx}^{(2)} & \chi_{zyy}^{(2)} & \chi_{zzz}^{(2)} & 0 & 0 & 0 \end{pmatrix} \begin{bmatrix} E_x^{\omega} E_x^{\omega} \\ E_y^{\omega} E_y^{\omega} \\ E_z^{\omega} E_z^{\omega} \\ 2E_y^{\omega} E_z^{\omega} \\ 2E_x^{\omega} E_z^{\omega} \\ 2E_x^{\omega} E_y^{\omega} \end{bmatrix}$$
$$= 2\varepsilon_0 \begin{pmatrix} 0 & 0 & 0 & 0 & d_{31} & -d_{22} \\ -d_{22} & d_{22} & 0 & d_{31} & 0 & 0 \\ d_{31} & d_{31} & d_{33} & 0 & 0 & 0 \end{pmatrix} \begin{bmatrix} E_x^{\omega} E_x^{\omega} \\ E_y^{\omega} E_y^{\omega} \\ E_z^{\omega} E_z^{\omega} \\ 2E_y^{\omega} E_z^{\omega} \\ 2E_x^{\omega} E_z^{\omega} \\ 2E_x^{\omega} E_y^{\omega} \end{bmatrix} \quad \text{(S1)}$$

Here, $d_{22}$, $d_{31}$, and $d_{33}$ are the independent nonlinear coefficients of LN. We consider a z-cut LN excited under normal excitation, with the FW propagating along the z direction and the generated SH signal collected in transmission along the same direction. For a normally incident plane wave, the electric field is transverse to the propagation direction. Therefore, to derive the CP-basis selection rule relevant to the normally transmitted SH field, we restrict the following analysis to the transverse nonlinear polarization components.

Under this transverse-field treatment, the nonlinear response of z-cut LN reduces to

$$\begin{aligned} P_x^{2\omega} &= 2\chi_{xxy}^{(2)} E_x^{\omega} E_y^{\omega} \\ P_y^{2\omega} &= \chi_{yxx}^{(2)} E_x^{\omega} E_x^{\omega} + \chi_{yyy}^{(2)} E_y^{\omega} E_y^{\omega} \end{aligned} \quad \text{(S2)}$$

where the nonzero transverse tensor elements are

$$\begin{aligned} \chi_{xxy}^{(2)} &= \chi_{yxx}^{(2)} = -d_{22} \\ \chi_{yyy}^{(2)} &= d_{22} \end{aligned} \quad \text{(S3)}$$

To reveal the spin-dependent nonlinear selection rule, we transform the transverse nonlinear polarization from the linear-polarization basis to the circular-polarization basis:

$$P^{2\omega}=\sum_{l_1}P_{l_1}^{2\omega}=\sum_{l_1}\varepsilon_0\sum_{l_2,l_3}\chi_{l_1l_2l_3}^{(2)}E_{l_2}^{\omega}E_{l_3}^{\omega},\quad \{l_1,l_2,l_3\}\in\{x,y\}$$
$$=\sum_{c_1}P_{c_1}^{2\omega}=\sum_{c_1}\varepsilon_0\sum_{c_2,c_3}\chi_{c_1c_2c_3}^{(2)}E_{c_2}^{\omega}E_{c_3}^{\omega},\ \{c_1,c_2,c_3\}\in\{L,R\} \tag{S4}$$

Here, $l_1$, $l_2$, and $l_3$ denote Cartesian components x and y, while $c_1$, $c_2$, and $c_3$ denote circular-polarization components L and R. Throughout this section, L and R are defined for waves propagating along the z direction of the transmissive geometry considered in this work. The corresponding circular-polarization unit vectors are defined as

$$\mathbf{e}_L=\frac{1}{\sqrt{2}}\left(\mathbf{e}_x+i\mathbf{e}_y\right),$$
$$\mathbf{e}_R=\frac{1}{\sqrt{2}}\left(\mathbf{e}_x-i\mathbf{e}_y\right) \tag{S5}$$

Accordingly, the Cartesian electric-field components can be expressed in terms of the circular-polarization components as

$$\begin{pmatrix}E_x\\E_y\end{pmatrix}=M\begin{pmatrix}E_L\\E_R\end{pmatrix} \tag{S6}$$

where

$$T_{cl}=\frac{1}{\sqrt{2}}\begin{bmatrix}1 & i\\ 1 & -i\end{bmatrix} \tag{S7}$$

and the inverse matrix is given by

$$T_{lc}=T^{-1}=\frac{1}{\sqrt{2}}\begin{bmatrix}1 & 1\\ -i & i\end{bmatrix} \tag{S8}$$

The circular components of the nonlinear polarization are obtained by projecting the Cartesian polarization onto the CP basis:

$$P_{c_1}^{2\omega}=\sum_{l_1}T_{cl}P_{l_1}^{2\omega}=\varepsilon_0\sum_{l_1,l_2,l_3}T_{cl}\chi_{l_1l_2l_3}^{(2)}E_{l_2}^{\omega}E_{l_3}^{\omega} \tag{S9}$$

The Cartesian electric-field components are written in terms of the CP components as

$$E_{l_2}^{\omega}=\sum_{c_2}T_{lc}E_{c_2}^{\omega} \tag{S10}$$

Substituting the field transformation into the Cartesian nonlinear polarization and then applying the CP projection, we obtain the nonlinear polarization on the CP basis:

$$P_{c_1}^{2\omega}=\sum_{l_1}T_{cl}P_{l_1}^{2\omega}=\varepsilon_0\sum_{c_2,c_3}\sum_{l_1,l_2,l_3}T_{cl}\chi_{l_1l_2l_3}^{(2)}T_{lc}E_{c_2}^{\omega}T_{lc}E_{c_3}^{\omega} \tag{S11}$$

By comparing this expression with the CP-basis definition of nonlinear polarization, the second-order susceptibility tensor in the circular-polarization basis is obtained as

$$\chi^{(2)}_{c_1c_2c_3} = \sum_{l_1,l_2,l_3} T_{cl}\chi^{(2)}_{l_1l_2l_3}T_{lc}T_{lc} \tag{S12}$$

The first index corresponds to the generated SH polarization, while the second and third indices correspond to the two FW fields involved in the SHG process. Expanding this transformation gives the six independent circular-basis tensor components:

$$\begin{aligned}
\chi^{(2)}_{RRR} &= \frac{\sqrt{2}}{4}[\chi^{(2)}_{xxx} - \chi^{(2)}_{xyy} + 2\chi^{(2)}_{yxy} + i(\chi^{(2)}_{yxx} - \chi^{(2)}_{yyy} - 2\chi^{(2)}_{xxy})] = 0 \\
\chi^{(2)}_{LLL} &= \frac{\sqrt{2}}{4}[\chi^{(2)}_{xxx} - \chi^{(2)}_{xyy} + 2\chi^{(2)}_{yxy} - i(\chi^{(2)}_{yxx} - \chi^{(2)}_{yyy} - 2\chi^{(2)}_{xxy})] = 0 \\
\chi^{(2)}_{RLL} &= \frac{\sqrt{2}}{4}[\chi^{(2)}_{xxx} - \chi^{(2)}_{xyy} - 2\chi^{(2)}_{yxy} + i(\chi^{(2)}_{yxx} - \chi^{(2)}_{yyy} + 2\chi^{(2)}_{xxy})] = -2\sqrt{2}id_{22} \\
\chi^{(2)}_{LRR} &= \frac{\sqrt{2}}{4}[\chi^{(2)}_{xxx} - \chi^{(2)}_{xyy} - 2\chi^{(2)}_{yxy} - i(\chi^{(2)}_{yxx} - \chi^{(2)}_{yyy} + 2\chi^{(2)}_{xxy})] = 2\sqrt{2}id_{22} \\
\chi^{(2)}_{RLR} &= \frac{\sqrt{2}}{4}[\chi^{(2)}_{xxx} + \chi^{(2)}_{xyy} + i(\chi^{(2)}_{yxx} + \chi^{(2)}_{yyy})] = 0 \\
\chi^{(2)}_{LLR} &= \frac{\sqrt{2}}{4}[\chi^{(2)}_{xxx} + \chi^{(2)}_{xyy} - i(\chi^{(2)}_{yxx} + \chi^{(2)}_{yyy})] = 0
\end{aligned} \tag{S13}$$

Substituting the z-cut LN tensor elements into the above expressions yields

$$\begin{aligned}
\chi^{(2)}_{RLL} &= -2\sqrt{2}id_{22} \\
\chi^{(2)}_{LRR} &= 2\sqrt{2}id_{22}
\end{aligned} \tag{S14}$$

while

$$\chi^{(2)}_{RRR} = \chi^{(2)}_{LLL} = \chi^{(2)}_{RLR} = \chi^{(2)}_{LLR} = 0 \tag{S15}$$

Therefore, the nonlinear polarization in the CP basis can be written as

$$\begin{aligned}
P^{2\omega}_{R} &= \chi^{(2)}_{RLL}E^{\omega}_{L}E^{\omega}_{L} \\
P^{2\omega}_{L} &= \chi^{(2)}_{LRR}E^{\omega}_{R}E^{\omega}_{R}
\end{aligned} \tag{S16}$$

These expressions show that, in the circular-polarization basis defined for waves propagating along the z direction, a left-handed circularly polarized FW generates a right-handed circularly polarized SH field, while a right-handed circularly polarized FW generates a left-handed circularly polarized SH field. The two allowed cross-circular SH channels have the same amplitude and differ by a relative phase of $\pi$. Thus, the z-cut LN layer acts as a spin-selective SH source with only two nonzero circular-basis nonlinear channels, namely $\chi^{(2)}_{RLL}$ and $\chi^{(2)}_{LRR}$.

### *S1.2 Experimental Verification of the FW-SH Polarization Mapping*

To experimentally verify the circular-polarization selection rule derived above, we measured the second-harmonic polarization generated from a bare z-cut LN substrate. The substrate has a thickness of 300 nm, identical to that used in the metasurface samples. Circularly polarized fundamental-wave (FW) excitation was provided by a femtosecond fiber laser (ARCHEv3, FYLA Laser) with a central wavelength of 1550 nm, a pulse duration of approximately 200 fs, and a repetition rate of 65 MHz. The measured FW spectrum and the generated SH spectrum are shown in Fig. S1(A) and Fig. S1(B), respectively.

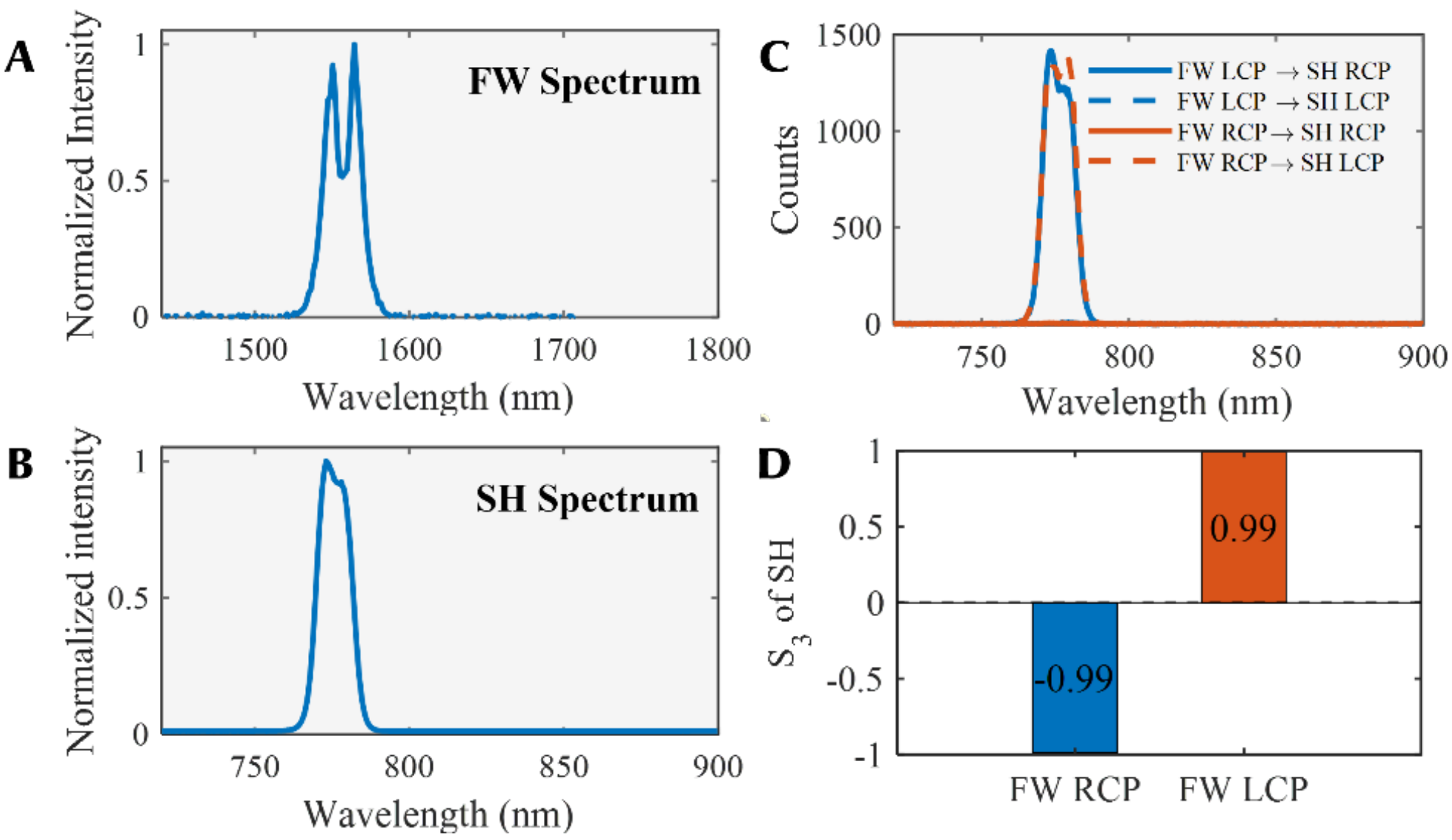

**Fig. S1. Experimental verification of FW and SH polarization mapping in a z-cut LN substrate.** (A) Spectrum of the femtosecond (fs) laser pump. (B) Measured SH spectrum. (C) SH spectra resolved in the circular polarization (CP) basis. (D) Polarization mapping between FW and SH, where the horizontal axis indicates the FW polarization state and the vertical axis shows the Stokes parameter $S_3$ of the SH signal.

The CP-resolved SH spectra in Fig. S1(C) shows that, under circularly polarized FW excitation, the generated SH signal is dominated by the circular-polarization channel opposite to that of the incident FW. This polarization mapping is further confirmed by the measured SH Stokes parameter $S_3$, as summarized in Fig. S1(D). The SH $S_3$ value is close to $-1$ under right-handed circularly polarized FW excitation and close to $+1$ under left-handed circularly polarized FW excitation, confirming the cross-circular FW–SH polarization mapping predicted by the tensor derivation.

## Section S2. Comparison Between Nonlinear Simulation and Linear Simulation Approximation

The refractive indices used in the simulations were set as follows. For $TiO_2$, $n$ = 2.3236 at 1550 nm and $n$ = 2.3688 + 1 × $10^{-6}$ $i$ at 775 nm. For LN, the ordinary refractive index is $n_o$ = 2.211 at 1550 nm and $n_o$ = 2.260 at 775 nm, while the extraordinary refractive index is $n_e$ = 2.138 at 1550 nm and $n_e$ = 2.180 at 775 nm. The refractive index of $SiO_2$ was taken as $n$ = 1.45.

As schematically illustrated in Fig. S2(A), we compare the two-step nonlinear simulation (NLS) with an equivalent one-step linear simulation (LS) model for the periodic unit cell. In the NLS model, the normally incident RCP FW first induces second-order nonlinear polarization inside the z-cut LN layer, and the generated SH field then propagates upward and is modulated by the $TiO_2$ nanopillar. When the periodic nanopillar array does not induce a pronounced resonant field redistribution inside the LN, the FW field distribution is only weakly perturbed, so the SH polarization remains primarily determined by the intrinsic $\chi^{(2)}$ response of z-cut LN and follows the cross-circular polarization mapping, described in Section S1. Under this condition, harmonic generation and wavefront manipulation can be treated as two largely independent processes. In this case, the nonlinear process can be approximated by an equivalent LS model, in which an LCP SH wave is directly launched from the LN layer and subsequently modulated by the nanopillar. In both models, the output SH response (amplitude and phase) is evaluated in the RCP component at port 2.

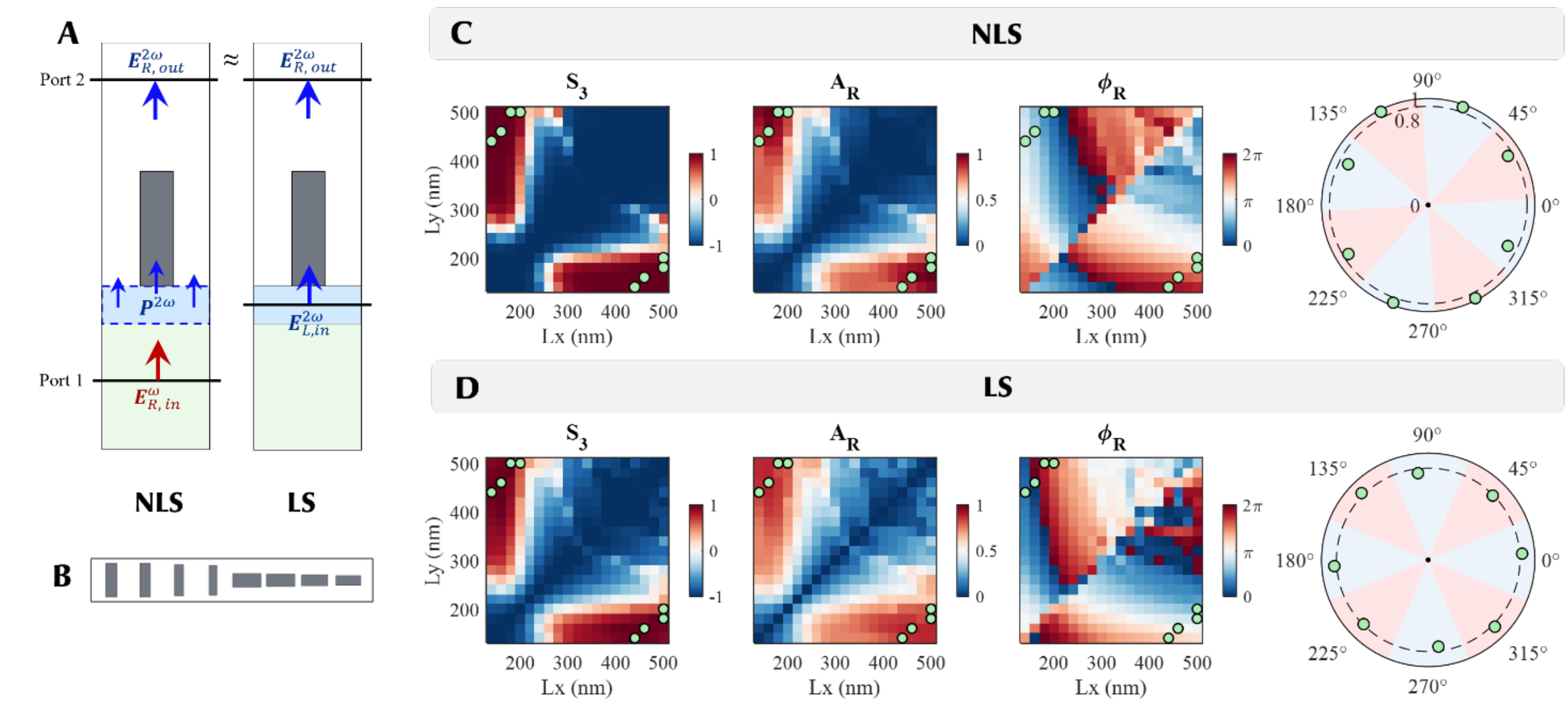

**Fig. S2. Comparison between nonlinear and linear simulations (NLS and LS) of the metasurface unit cells.** (A) Schematic illustration of the NLS and LS models. (B) Cross-sectional views of the eight selected meta-atoms identified from the NLS results. (C) Simulated maps of polarization purity ($S_3$), amplitude (A), and phase (φ) as functions of the lateral dimensions (Lx, Ly) in the NLS case. (D) Corresponding maps obtained from LS. The selected meta-atoms (marked in Fig. S2C), satisfying $S_3$ > 0.92 and A > 0.8, are projected onto the LS maps and exhibit similarly high performance in terms of both half-wave plate (HWP) functionality and phase coverage. This agreement indicates that the designed structures operate in a non-resonant regime, allowing the SH field in the LN layer to be approximated as nearly ideal circularly polarized light.

| Index | $L_x$ (nm) | $L_y$ (nm) | $S_3$ | $A_R$ | $\phi_R$ (deg) |
|---|---|---|---|---|---|
| 1 | 200 | 500 | 0.929 | 0.879 | 31.6 |
| 2 | 180 | 500 | 1.000 | 0.974 | 70.5 |
| 3 | 160 | 460 | 1.000 | 0.982 | 116.9 |
| 4 | 140 | 440 | 0.993 | 0.842 | 152.9 |
| 5 | 500 | 200 | 0.929 | 0.879 | 211.6 |
| 6 | 500 | 180 | 1.000 | 0.974 | 250.3 |
| 7 | 460 | 160 | 1.000 | 0.982 | 296.8 |
| 8 | 440 | 140 | 0.993 | 0.842 | 332.8 |

**Table S1. Geometrical parameters and NLS results of the selected eight meta-atoms.**

To verify this approximation, we first selected eight nanopillar geometries from the NLS results using the criteria ($S_3 > 0.92$) and ($A_R > 0.80$), and a nearly uniform coverage of $\phi_R$ over the full 0–$2\pi$ range. Here, $A_R$ denotes the amplitude of the RCP component of the output SH field collected at port 2, normalized to the SH amplitude generated from the bare LN substrate. The geometrical parameters and corresponding NLS results of the selected nanopillars are summarized in Table S1. These selected geometries were then projected onto the LS maps, where they show similar performance, with an average $S_3 = 0.91$ and an average RCP-channel transmission amplitude $A_R = 0.86$. This agreement indicates that the LS model provides a more computationally efficient approach for unit-cell selection and also enables experimental pre-characterization of the metasurface performance using a linear light source at the SH wavelength, as further verified by the holographic comparison in Fig. S3.

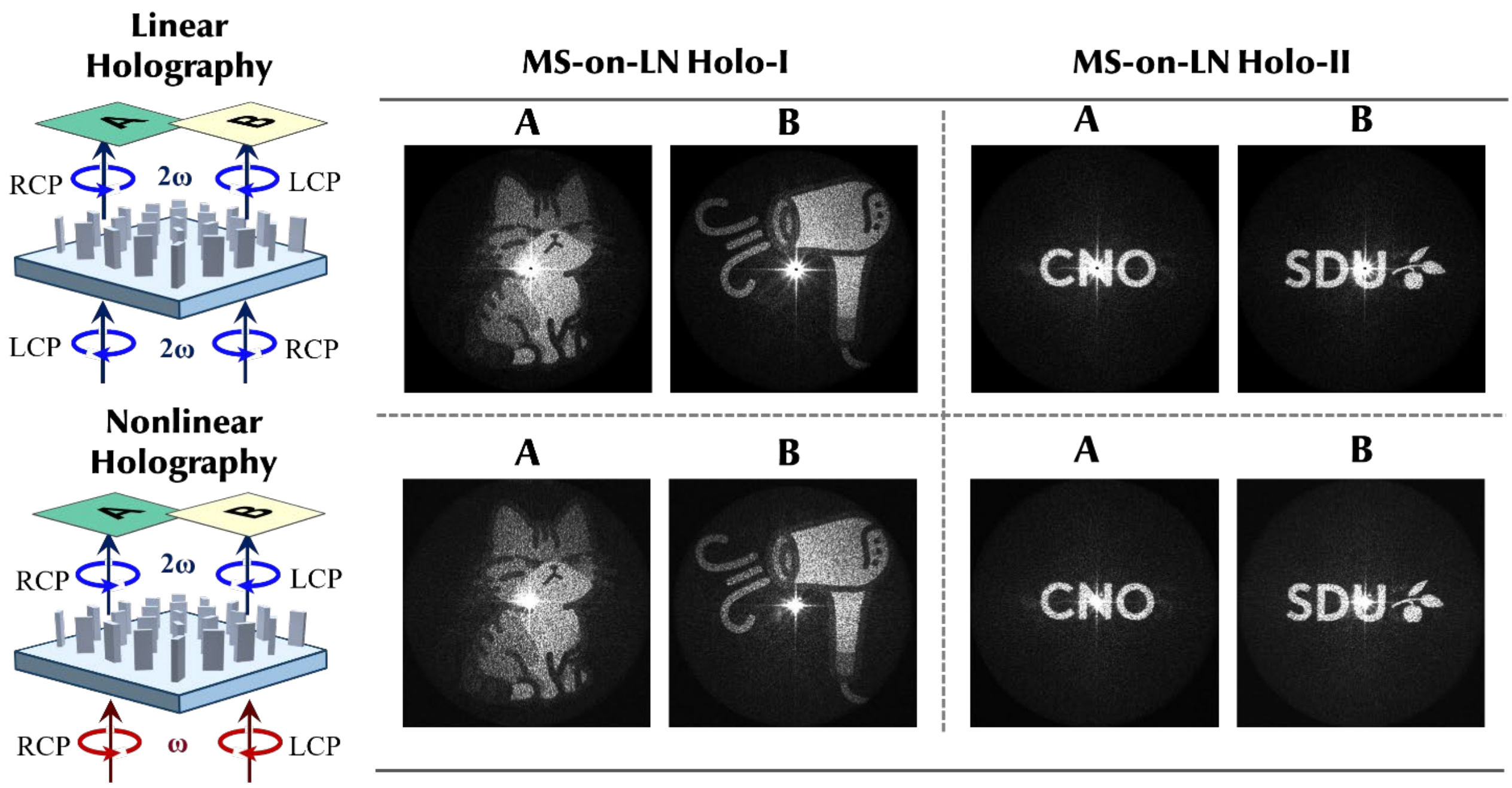

**Fig. S3. Experimental comparison of holographic image generation under linear and nonlinear excitations for two designs (MS-on-LN Holo-I and MS-on-LN Holo-II).**

### Section S3. Device Fabrication and SEM Images

The fabrication of the metasurface was carried out using electron-beam lithography (EBL), thin-film deposition, and inductively coupled plasma (ICP) etching processes. The structures were fabricated on a commercially sourced z-cut LN bonded to a 2.5 μm-thick silicon dioxide ($SiO_2$) buffer layer atop a 500 μm-thick quartz substrate (NanoLN, Jinan Jingzheng Electronics Co.). Prior to fabrication, the substrate was sequentially cleaned in acetone and ethanol, followed by rinsing in deionized water and drying with a nitrogen gun. A 800-nm-thick $TiO_2$ thin film was then deposited using an optical thin-film deposition system, and the thickness was verified by a surface profilometer. For the lithography process, a layer of ZEP520A electron-beam resist (thickness of ~170 nm) was spin-coated onto the $TiO_2$ surface, followed by a post-application bake. Electron-beam exposure was performed using a beam current of 2 nA. After exposure, the sample was developed for 90 s and then immersed in a fixer for 30 s. A chromium (Cr) mask layer with a thickness of 50 nm was subsequently deposited. A lift-off process was carried out by immersing the sample in heated acetone, resulting in the transfer of the patterned Cr mask onto the substrate. The $TiO_2$ layer was then etched to a depth of 800 nm using an inductively coupled plasma (ICP) etching system. Finally, the residual Cr mask was removed using a chromium etchant solution, leaving the desired $TiO_2$ nanostructures on the substrate.

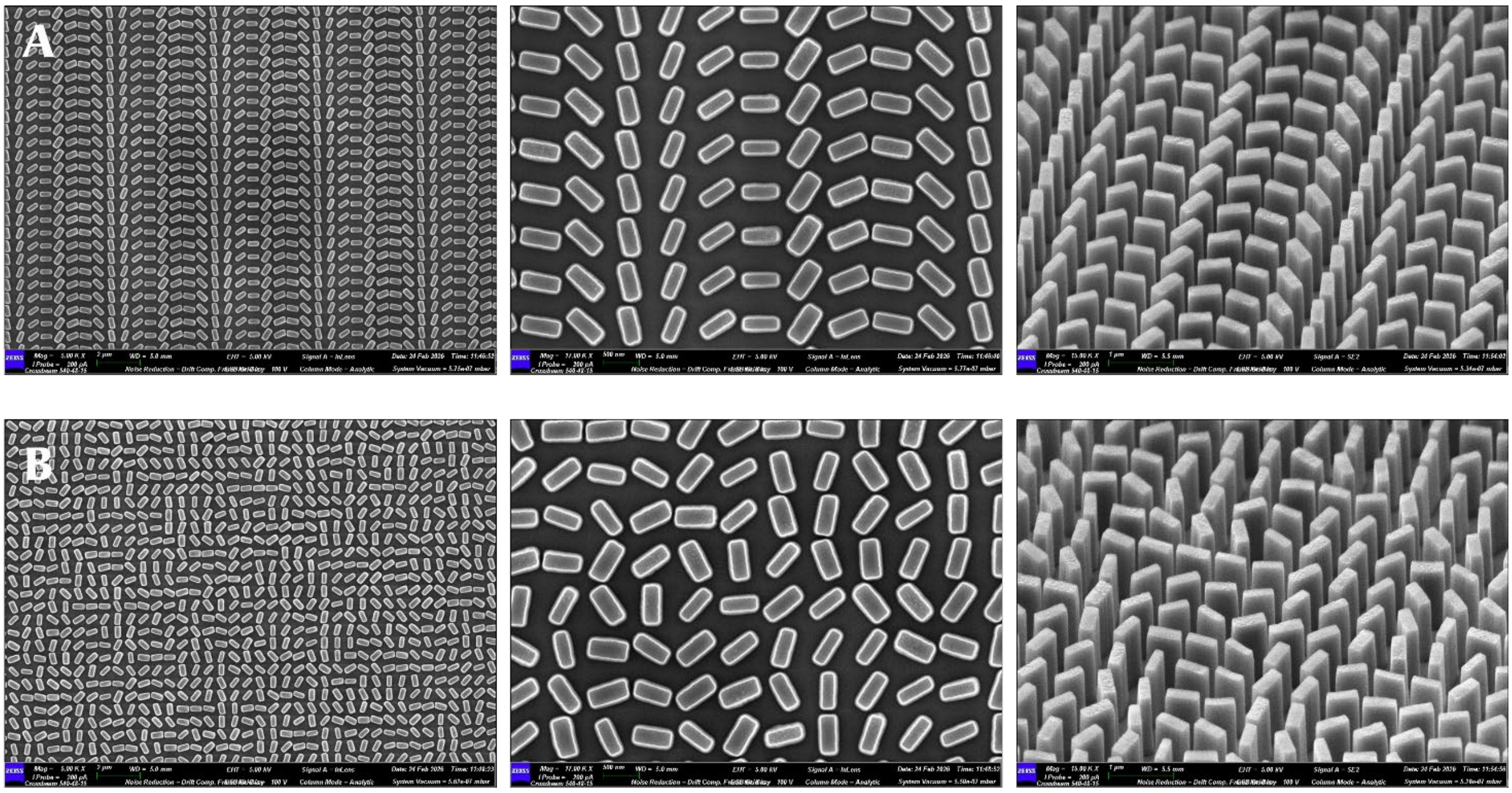

**Fig. S4. SEM images of the (A) MS-on-LN grating and (B) MS-on-LN Holo-I.**

## Section S4. Optical Setup for Linear and Nonlinear Characterization

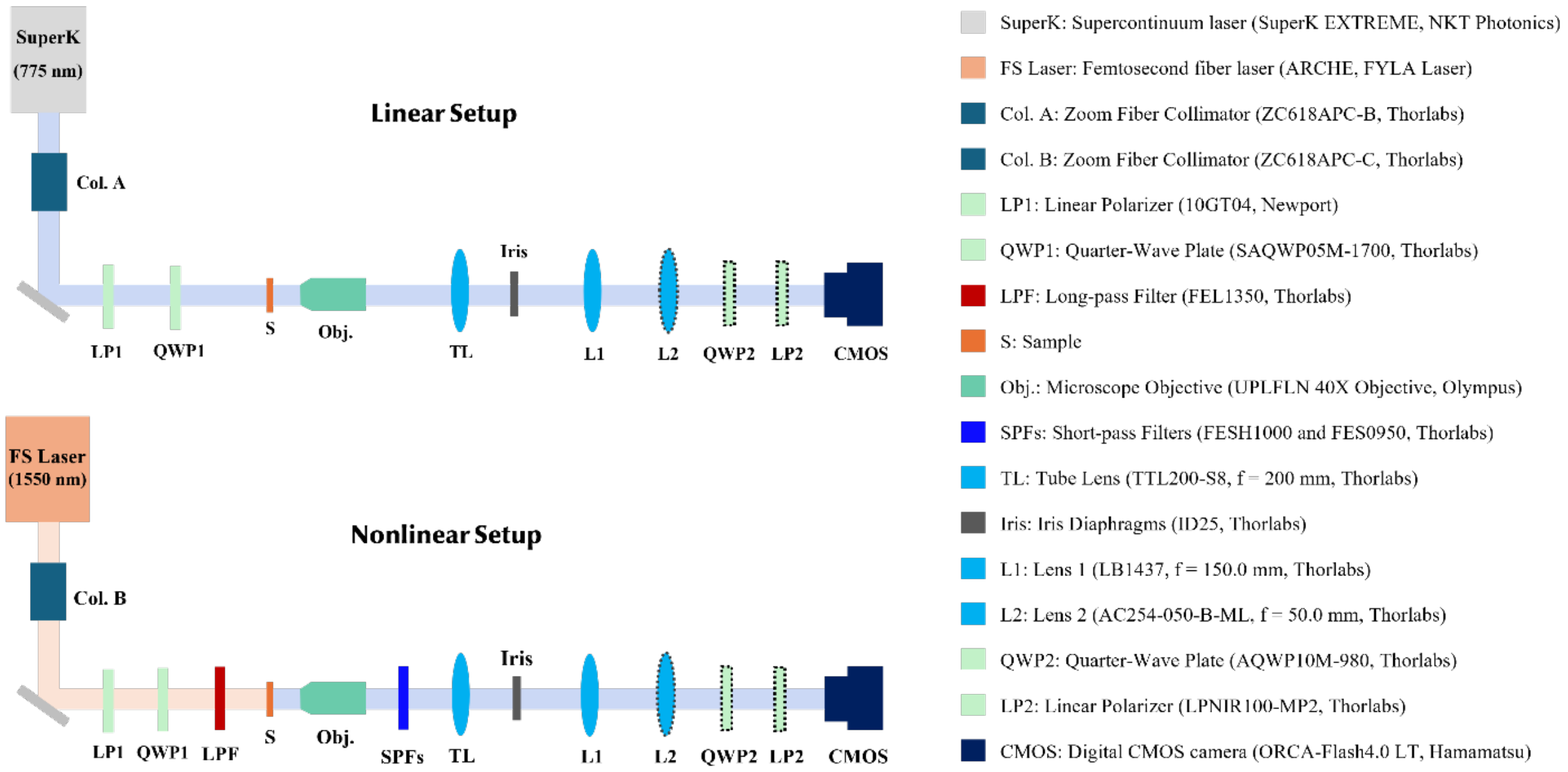

**Fig. S5. Experimental setup for linear and nonlinear characterization of the MS-on-LN samples. All optical components are listed in the right panel.** The linear setup operates at the second-harmonic (SH) wavelength (775 nm) using a filtered supercontinuum laser source, enabling pre-characterization of the metasurface response, including beam steering and holography, under linear conditions. The nonlinear setup employs a femtosecond (fs) laser with a central wavelength of 1550 nm. A long-pass filter (LPF) and short-pass filters (SPFs) are used to suppress the fundamental wave and isolate the SH signal generated from the sample.

**Section S5. Simulated Spin-Dependent Diffraction Performance of the MS-on-LN Grating**

To evaluate the diffraction performance of the designed MS-on-LN grating, we performed nonlinear simulations under LCP and RCP FW excitation. Figures S6(A, B) show the simulated spatial distributions of the $E_L^{2\omega}$ and $E_R^{2\omega}$ components in the *x*-*z* plane. The generated SH fields are predominantly concentrated in the designed co-polarized output channels with respect to the FW excitation and are deflected along different propagation directions, consistent with the spin-dependent phase gradients encoded in the supercell. The corresponding order-resolved relative diffraction efficiencies are shown in Figs. S6(C, D), where the intensity of each order is normalized to the 0th-order co-polarized SH intensity generated from the bare LN substrate. For LCP FW excitation, the SH signal is predominantly directed into the diffraction order $m = -1$ in the L → L channel, with a relative efficiency of 0.881. For RCP FW excitation, the SH signal is mainly directed into the diffraction order $m = +2$ in the R → R channel, with a relative efficiency of 0.747. The numerical procedure used to obtain these order-resolved efficiencies is illustrated in Fig. S6(E), where the SH electric field is extracted from an output plane above the supercell and decomposed into diffraction orders by Fourier transformation.

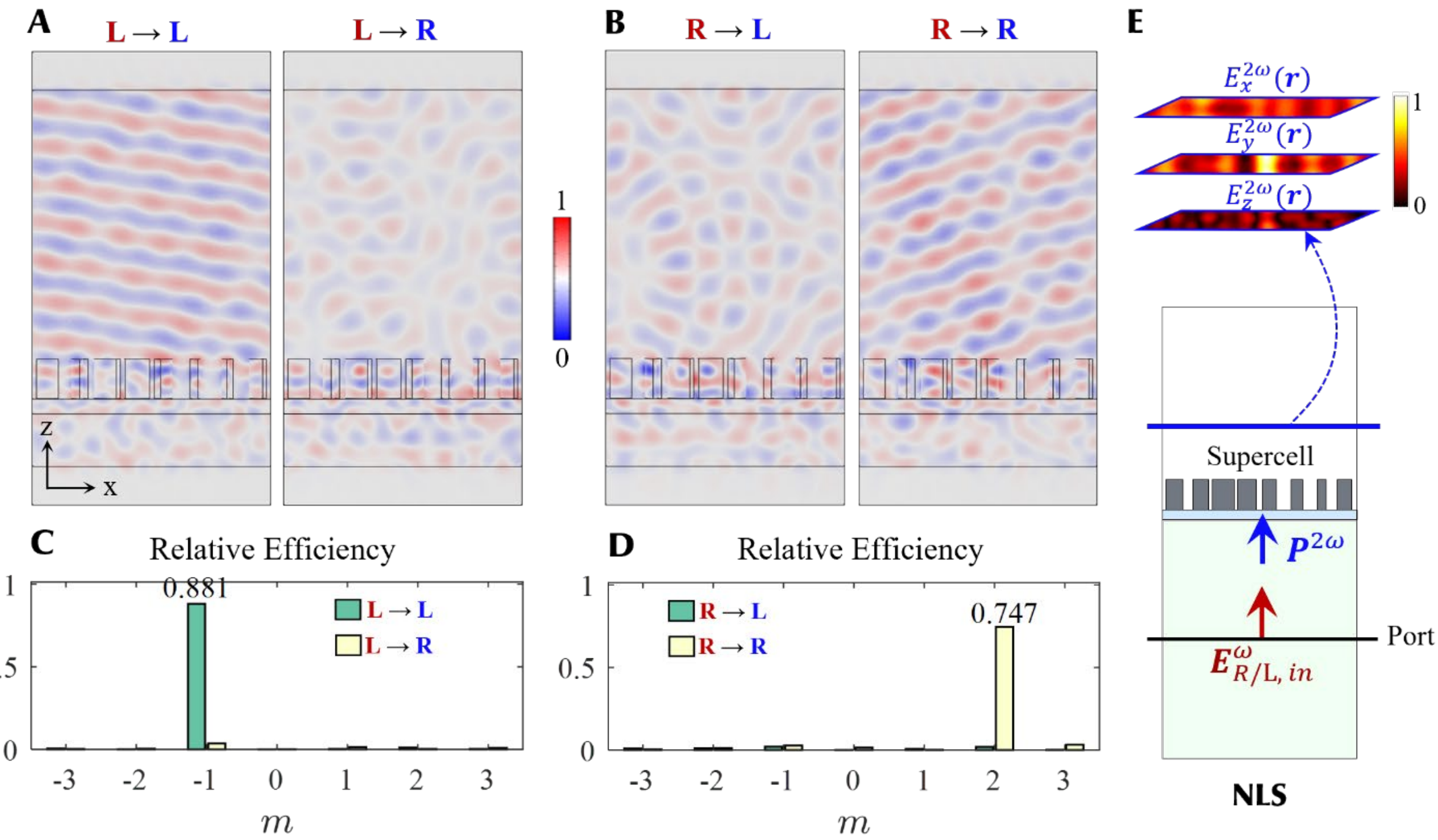

**Fig. S6. Simulated diffraction performance of the MS-on-LN grating.** (A, B) Simulated spatial distributions of the $E_L^{2\omega}$ and $E_R^{2\omega}$ components of the generated SH field from the MS-on-LN grating supercell under LCP and RCP FW excitation, respectively. (C, D) Relative diffraction efficiencies of the generated SH signals at different diffraction orders under LCP and RCP FW excitation, respectively. The efficiencies are normalized to the 0th-order SH intensity generated from the bare LN substrate. (E) Schematic of the nonlinear simulation and post-processing procedure used to obtain the diffraction-efficiency results in Fig. S6 (C, D). The SH electric field is extracted from an output plane above the supercell and then Fourier-transformed for polarization- and order-resolved far-field analysis.

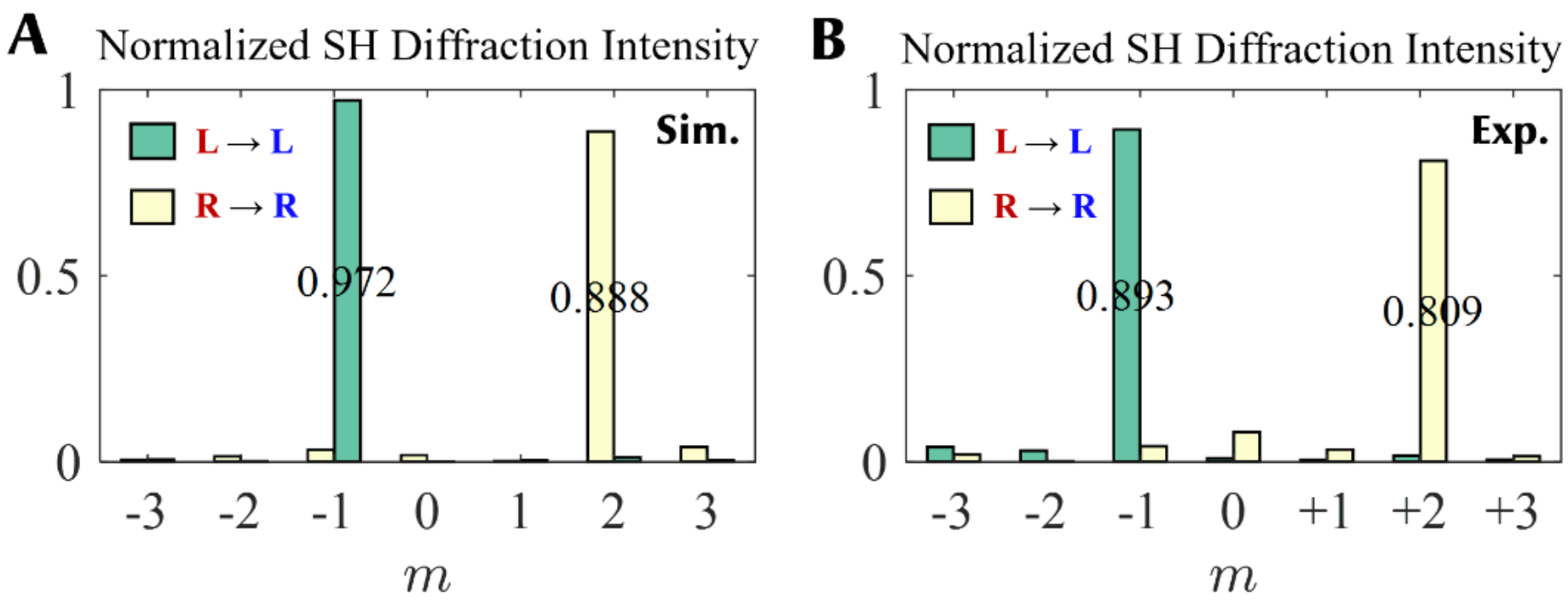

**Fig. S7. Simulated and measured normalized diffraction-order intensity distribution of the MS-on-LN grating.** For each input polarization, the diffraction-order intensities are normalized to the summed SH intensity over the displayed orders. The designed $m = -1$ and $m = +2$ orders show normalized intensities of 0.972 and 0.888 in simulation, and 0.893 and 0.809 in experiment, respectively. The strong concentration of SH power in the designed diffraction orders confirms efficient spin-dependent routing of the generated SH signals.

To further evaluate the diffraction contrast of the MS-on-LN grating, we compare the simulated and measured SH intensity distributions across the displayed diffraction orders, as shown in Fig. S7. This analysis shows the relative brightness of the target diffraction order compared with the other diffraction orders. In simulation, the designed $m = -1$ and $m = +2$ orders exhibit normalized diffraction-order intensities of 0.972 and 0.888 in simulation, and 0.893 and 0.809 in experiment, respectively. These results confirm that the generated SH signal is primarily concentrated in the designed diffraction orders, with good consistency between simulation and experiment.

## Section S6. Hologram Design and Workflow and Reconstruction Analysis

In this section, we show the hologram design process and the influence of several practical factors on the reconstructed image quality. As shown in Fig. S9, the target image is first pixelated and then reconstructed after phase retrieval, 8-level phase quantization, and finally replacing the ideal phase-only pixels with the simulated amplitude and phase responses of the selected metasurface unit cells. The comparison demonstrates that the main image features are preserved throughout the design process, while the reconstruction becomes dimmer after discretizing the continuous phase into eight levels. The influence of phase quantization is further examined in Fig. S10, where increasing the number of phase levels generally improves the reconstructed image quality. In this work, an 8-level phase library is adopted as a proof-of-concept implementation, balancing design simplicity and reconstruction performance.

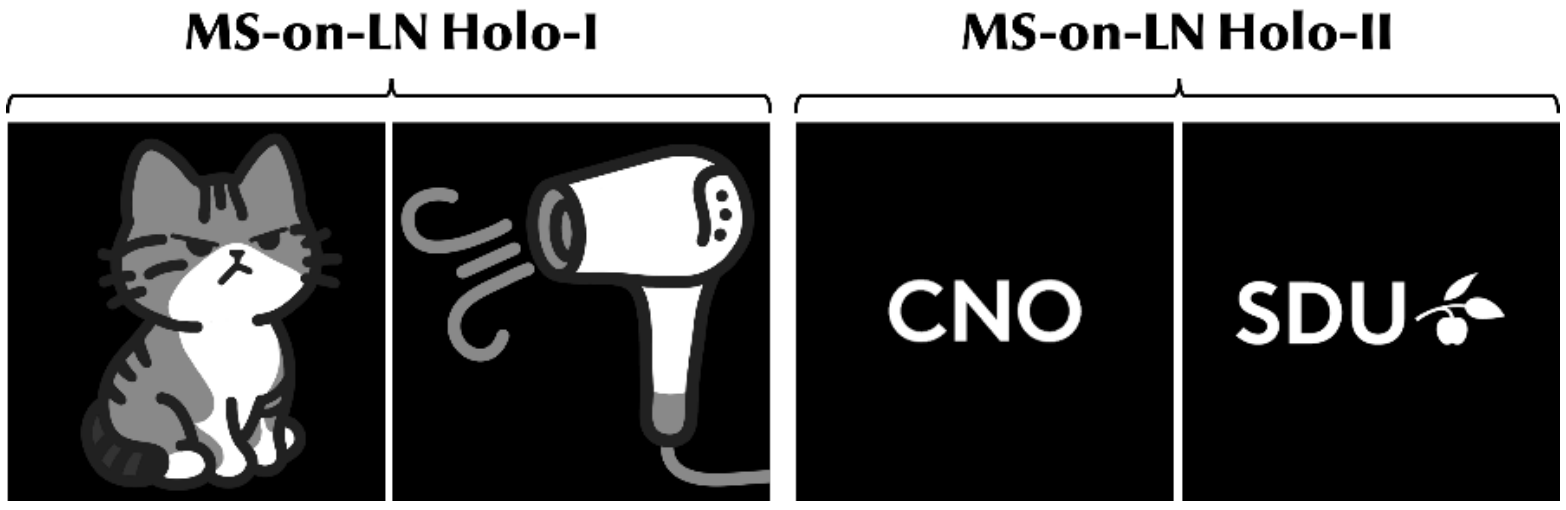

**Fig. S8. Target holographic images encoded in MS-on-LN Holo-I and MS-on-LN Holo-II.**

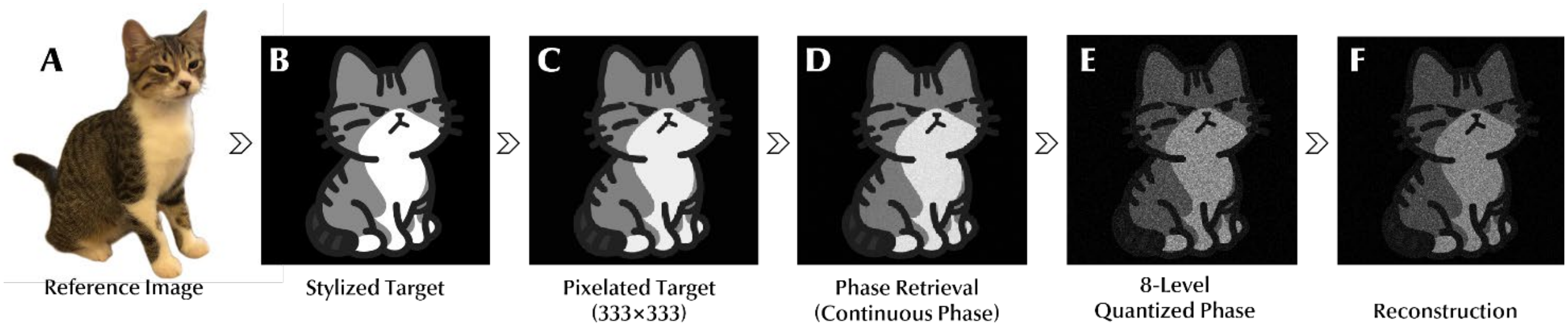

**Fig. S9. Evolution of holographic image quality throughout the design process, including pixelation, phase retrieval, and quantization.** (A) Reference image. (B) Stylized target image. (C) Pixelated target (333 × 333 unit cells). (D) Continuous phase distribution retrieved using a stochastic gradient descent (SGD) algorithm and reconstructed via Fourier transform. (E) Reconstruction using 8-level phase quantization. (F) Reconstruction using the discrete phase and amplitude responses of the designed meta-atoms.

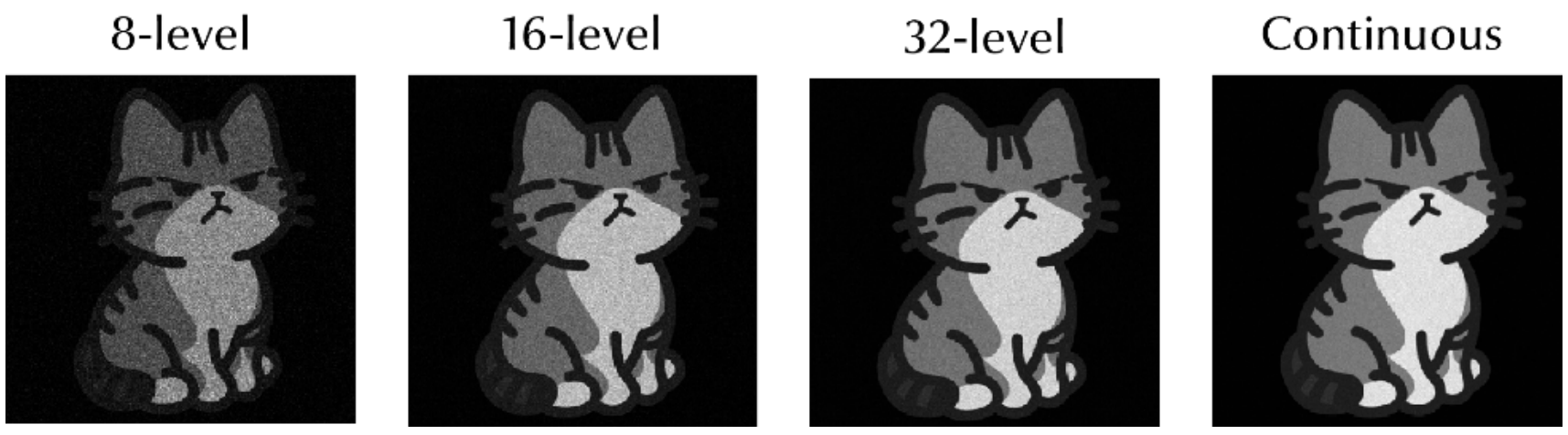

**Fig. S10. Effect of phase quantization levels on holographic image reconstruction quality.**

We further investigate the influence of the incident Gaussian beam size on holographic reconstruction, as shown in Fig. S11, to provide a reference for choosing the beam size in the experiment. As the beam waist increases, the illumination over the metasurface becomes closer to a uniform plane wave, leading to improved image quality for both linear holography (LH) and nonlinear holography (NLH). The NLH reconstruction is more sensitive to the FW beam size because the generated SH intensity scales quadratically with the local FW intensity. Therefore, a larger FW beam is preferred for uniform hologram illumination, while a sufficient power density must still be maintained for detectable SH generation, leading to a practical trade-off in the experiment.

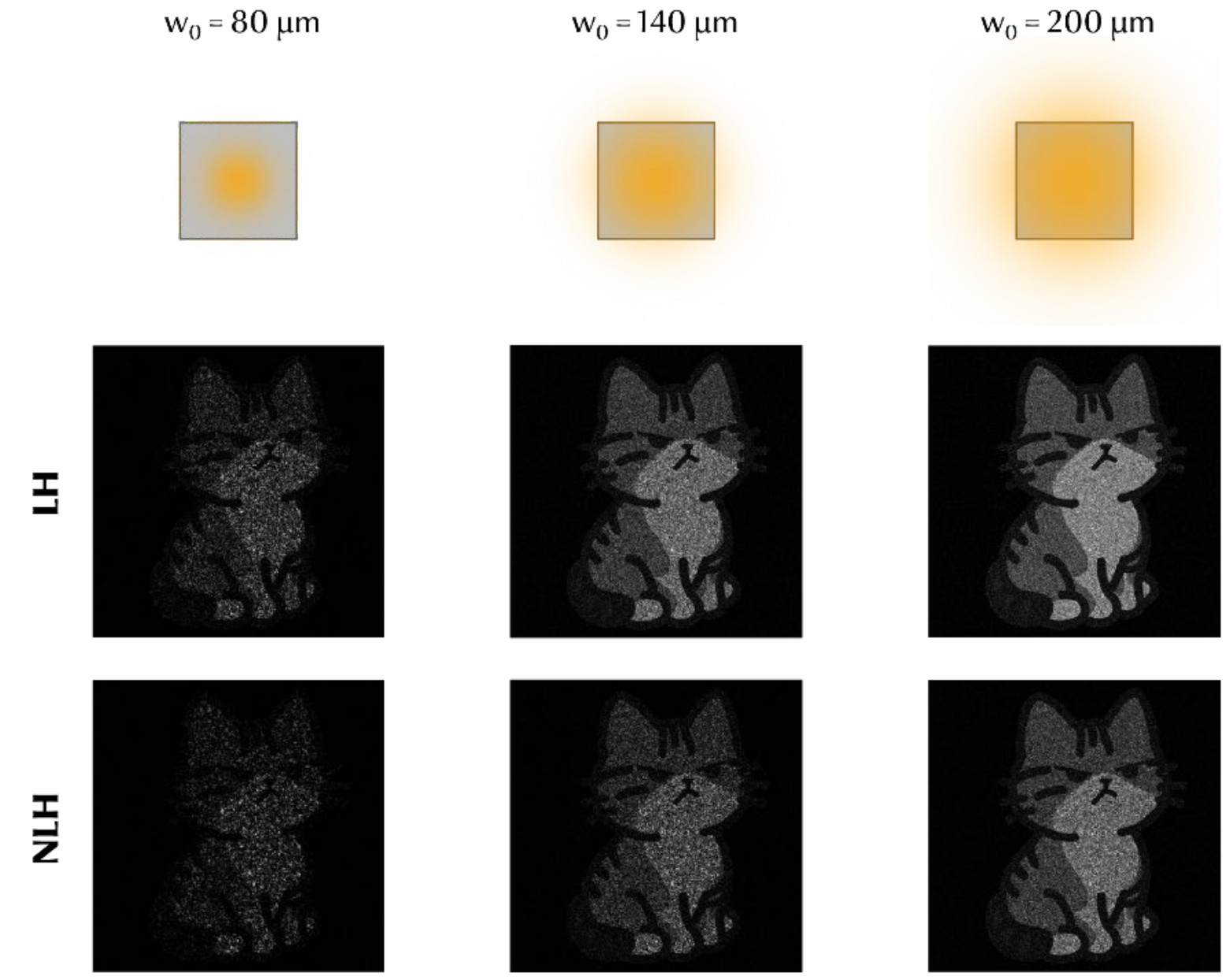

**Fig. S11. Effect of Gaussian beam size on holographic image reconstruction quality.** Simulated holographic images under different incident beam waists $w_0$. The top row illustrates the Gaussian beam profile relative to the metasurface area. The middle and bottom rows show the reconstructed images for linear holography (LH) and nonlinear holography (NLH), respectively. As the beam waist increases, the illumination becomes more uniform across the metasurface area, leading to improved image quality, particularly for NLH. The metasurface size is approximately 200 × 200 μm$^2$.

## Section S7. Supplementary Results of Spin-Multiplexed SH Holography

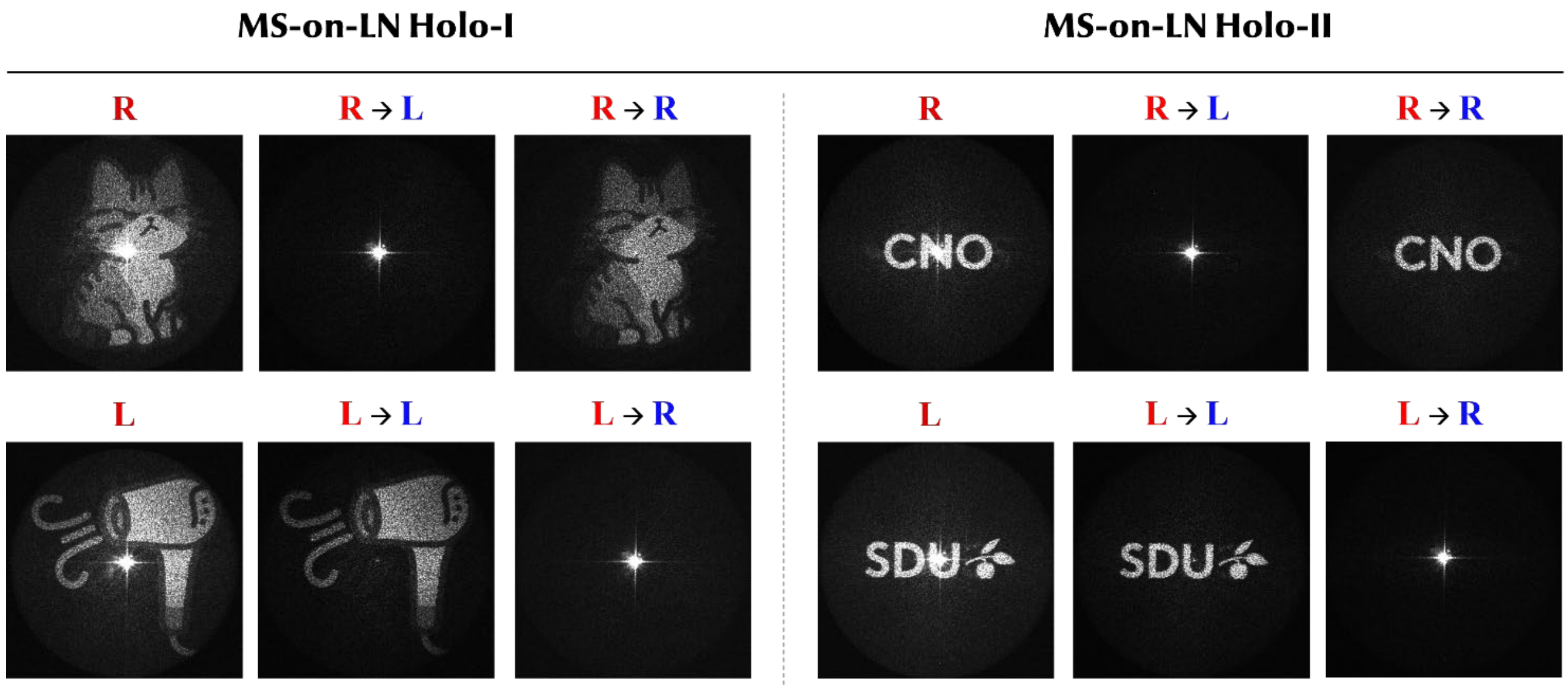

**Fig. S12. Experimental results of spin-multiplexed SH holography.** Reconstructed SH holographic images from MS-on-LN Holo-I and MS-on-LN Holo-II under orthogonal circularly polarized FW excitations. The polarization-resolved measurements confirm that the holographic images are reconstructed in the designed co-polarized SH channels, while the residual 0th-order signals are predominantly observed in the cross-polarized channels.

## S8 Thickness-Dependent SHG Efficiency in z-Cut LN

SHG efficiency is an important consideration for practical applications, as higher efficiency improves the signal-to-noise ratio and reduces the required pump power. Under the undepleted-pump approximation, i.e., in the limit of low conversion efficiency, the second-harmonic generation (SHG) efficiency varies with the interaction length $l$ as follows:

$$\eta_{SHG}(l) \propto \sin^2\left(\frac{\pi l}{2 l_c}\right) \tag{S17}$$

Here,

$$l_c = \frac{\lambda_\omega}{4\left|n_{2\omega} - n_\omega\right|} \tag{S18}$$

is the coherence length, where $\lambda_\omega$ is the FW wavelength, $n_\omega$ and $n_{2\omega}$ are the refractive indices of LN at the FW and SH wavelengths, respectively. In our case, the coherence length is approximately 7.9 μm, which is much larger than the 300-nm LN film thickness used in this work. The SHG efficiency reaches its first maximum when the interaction length equals $l_c$ and is estimated to be approximately 280 times higher than that for a 300-nm interaction length. This indicates that the SHG efficiency can be tailored by appropriately selecting the LN film thickness. The overall design principle remains unchanged, and if the modified LN thickness does not lead to the excitation of guided-mode resonances at the design wavelength, the metasurface design may even remain unchanged while providing the same wavefront-control functionality.